\documentclass[12pt]{article}
\usepackage{graphicx} % Required for inserting images
\usepackage[utf8]{inputenc}
\usepackage{geometry}

\usepackage{booktabs}
\usepackage{amssymb}
\usepackage{amsmath}
\usepackage{lscape}
\usepackage{bbm}
\usepackage{xcolor}
\usepackage{url}
\usepackage{footmisc}

\newtheorem{lemma}{Lemma}

\newtheorem{proposition}{Proposition}
\newtheorem{remark}{Remark}

\newtheorem{assumption}{Assumption}
\usepackage{setspace}
\usepackage{footmisc}
\usepackage[round]{natbib}
\usepackage[final]{hyperref} % adds hyperlinks inside the generated pdf file

\usepackage{enumerate}   
\usepackage{tikz}

\usepackage{accents}

\usepackage{chngcntr}
\usepackage{apptools}
\AtAppendix{\counterwithin{lemma}{section}}
\AtAppendix{\counterwithin{proposition}{section}}
\title{Team Disagreement and Productive Persuasion\thanks{I am grateful to Jonathan Bonham,  Renee Bowen, Carlo Cusumano, Simone Galperti, Étienne Gagnon, Nicola Gennaioli, Germán Gieczewski, Aram Grigoryan, Gleason Judd, Navin Kartik, Kris Ramsay, Denis Shishkin, Joel Sobel, Pietro Spini, Guido Tabellini, Davide Viviano, and UCSD seminar participants for helpful comments and suggestions.}}
\author{Giampaolo Bonomi\thanks{Princeton University. Email: bonomi@princeton.edu}}
\date{December 2025}

\begin{document}

\maketitle
\vspace{.5cm}
\begin{abstract}
We study how open disagreement influences team performance in a dynamic production game. Team members can hold different priors about the productivity of the available production technologies. Initial beliefs are common knowledge and updated based on observed production outcomes. We show that when only one technology is available, a player works harder early on when her coworkers are initially more pessimistic about the technology's productivity. Holding average team optimism constant, this force implies that a team's expected output increases in the degree of disagreement of its members. A manager with the task of forming two-member teams from a large workforce maximizes total expected output by matching coworkers' beliefs in a negative assortative way. When alternative, equally good, production technologies are available, a disagreeing team outperforms any like-minded team in terms of average output and team members' welfare. 
\vspace{.4cm}


\end{abstract}
\newpage
\section{Introduction}
\begin{quote}
\begin{singlespace}
\textit{``I don't feel that an atmosphere of debate and total disagreement and argument is such a bad thing. It makes for a vital and alive field.''
\begin{flushright} (Clifford Geertz)\end{flushright}}  
\end{singlespace}
\end{quote}

Conventional wisdom suggests that the interaction of people with different backgrounds and perspectives will often lead to socially desirable outcomes. Indeed, economists have advanced intuitive arguments linking diversity to successful problem-solving and improved decision-making, relying on the idea that different perspectives and capabilities naturally serve as complements, enriching and refining each other \citep[e.g.,][]{Hong2001,Hong2004,Page}. There is much less focus on the productivity implications of open and persistent \textit{disagreement}, characterized by strongly conflicting views and interpretations of a problem. In this paper we address the following question: can disagreement in a team of innovators boost the team's output? When should we expect this to happen? We bring to light a force specific to disagreement: the incentive to persuade others through production breakthroughs. In short, disagreement increases output because optimistic agents work harder initially to convince pessimists, and this incentive outweighs the reduced effort of pessimists.

Seminal contributions to the economic literature have warned us about the perils of preference disagreement, shown to create impasse and inefficiencies in many domains of social interaction.\footnote{For instance, disagreement has been shown to impede decision-making and compromise economic outcomes in social choice \citep[e.g.][]{Arrow1951}, communication \citep[e.g.,][]{CS}, public finance and public good provision \citep[e.g,][]{AlesinaT1990,AlesinaLaF2005}.} Yet, history suggests a link between the conflict of worldviews and greater innovation. The development of the first iPhone was, anecdotally, a story of disagreement \citep{iPhone,Grant2021}. Steve Jobs initially thought the product would only appeal to a ``pocket protector'' crowd, and saw the project as a dead end. A team of hard-working engineers --- with Apple's design chief Jonathan Ive on their side ---  had a very different opinion. They believed that the touchscreen technology would represent a paradigm shift for the industry. Job's skepticism meant that the team needed to design a prototype so good that it would have been impossible for him not to change his mind.\footnote{According to \citet{Grant2021}: \textit{``Fadell and his engineers chipped away at the resistance by building early prototypes in secret, showing Jobs demos, and refining their designs.''}} Some argue that Job's open disagreement culture was the key to the company's success in those years \citep[e.g.,][]{Scott2017,Grant2021}.    

Beyond the iPhone anecdote, the power of disagreement has been recognized in many innovation-related contexts. Open disagreement between scientists has led to the production of more and better theories, resulting in the belief that scientific skepticism --- the tendency to challenge and falsify existing theories --- lies naturally at the heart of scientific progress \citep{Kuhn1962}. Peers' skepticism has typically motivated philosophers to design sophisticated arguments in favor of their worldview, to convince others to adopt it.\footnote{For instance, determined to convince skeptics about the existence of God, Anselm of Canterbury designed the first \textit{ontological argument}, a class of arguments that has fascinated philosophers for almost a thousand years (see \url{https://plato.stanford.edu/entries/ontological-arguments/}).} Artists have often found in partners' disagreement and competition of ideas a motivating force inspiring them to innovate and often reach success. \citet{Metallica} illustrate this point using the case study of the heavy metal band \textit{Metallica} and many of their historical producers (most notably, producer Bob Rock): \textit{``Both sides [Metallica and Rock] wanted to make the best record in the world but disagreed on what it
should be like and how it should be done. [...] the most constructive outcomes of conflicts realized when Metallica collaborated
with its partners but at the same time also competed with them. The partners first competed
to find the best idea, after which the best idea was further developed together.''} 

Outside the realm of intellectual and artistic debates, the \textit{space race} was perhaps one of the starkest examples of how the collision of conflicting beliefs about the ideal economy and society spurred technological investment, leading to remarkable innovations in an attempt to prove the superiority of one worldview over the other \citep{TechPolitik}.\footnote{According to Brian C. Odom, Nasa Chief Historian: \textit{``In the global South, you had a lot of countries becoming independent from former colonial powers. What system would they follow? Would they follow the U.S. liberal democracy or would they follow the Soviet example of communism? Kennedy saw the race to the moon as a way to demonstrate American technological power and the benefit of one system over another.''} See \url{https://www.space.com/space-race.html}.} The more recent \textit{standards wars} in the Tech industry --- battles for market dominance between incompatible technologies --- can be thought of in a similar fashion.\footnote{R\&D breakthroughs are recognized as typical ways of persuading the key players and winning such wars. This was the case, for instance, in the battle between the AC and DC technologies for the generation and distribution of power. See \citet{ShapiroVarian} for a discussion of  standard wars. }

The above examples share three common features: (i) the players involved hold extremely different beliefs about the promise of one or more production technologies and are aware of such disagreement; (ii) some players benefit from persuading others to change their minds or switch to their production approach; and (iii) persuasion occurs through the successful results of productive effort. 

This paper explores the productivity implications of disagreement in a simple class of games sharing the above three characteristics. A team of coworkers engage in a two-period production game. In each period, they simultaneously and independently choose levels and allocation of costly effort across one or more production technologies. At the end of each period, the team's output is realized, and payoffs --- increasing in such output --- accrue. We make the following key assumptions. First, coworkers are uncertain about the returns on effort of the available production technologies --- high or low --- and can start the game with different prior beliefs about these returns. Team members are aware of each others' priors, and whenever these priors differ, they think others are wrong: they \textit{disagree}.\footnote{Note that awareness of \textit{pior} disagreement does not contradict \citet{Aumann}'s ``agreeing to disagree'' result, which \textit{assumes} a common prior.} Second, the higher the effort invested in a technology, the more likely the technology is to produce a return (breakthrough) and reveal its value. Third, players observe past breakthroughs and update their beliefs accordingly.

We start by considering the case of a single available technology and define team optimism and team disagreement as, respectively, the average prior probability that the team members assign to the event that the technology yields high returns, and the average squared distance between the priors of any two team members. We find that each worker's first-period effort is increasing in the pessimism of her coworkers. Intuitively, when a team member is more optimistic than her coworkers, she is motivated to work harder early on, to obtain the high returns that persuade others to exert higher effort in future periods. If instead she is more pessimistic than her coworkers, she expects return arrivals to make the team, on average, more pessimistic and less productive in the future, and she is therefore induced to work less early on. Both forces are muted if all team members share the same view. Second, and related, we find that, holding team optimism constant, a team's expected output increases with the level of team disagreement. In particular, we show that team output can be decomposed into two additive terms, one only depending on team optimism, and the other directly proportional to team disagreement: when treated as inputs for team production, disagreement and optimism can be thought of as substitutes. This \textit{disagreement dividend} is entirely generated by how persuasion incentives influence players' first-period effort choices.

Despite the presence of a disagreement dividend, when only one technology is available, team expected output is maximized by a fully optimistic team. Hence, the one-technology results naturally lend themselves to contexts where team managers cannot arbitrarily ``choose'' the beliefs of team members but need to form teams optimally for a given belief composition of their pool of workers. In this spirit, we proceed by studying the problem faced by a manager who needs to allocate a continuum of disagreeing employees into two-member teams, in a way that maximizes the aggregate expected output of the workforce. Formally, we solve the optimal transport problem of distributing a continuum of priors into a continuum of prior-pairs, with the value of each pair defined as the equilibrium expected output attained in our production game by a two-member team with such initial beliefs. Building on our previous results, we show that, regardless of the prior beliefs of the team manager, there is a unique solution entailing negative sorting, with the most optimistic employees paired with the most pessimistic ones. Hence, the manager should deliberately form teams with extreme disagreement. In addition, under a symmetry condition, the total expected output of the workforce in the optimal matching increases with the disagreement of the total workforce.

Why is negative sorting optimal, and why does the value of the optimal transport problem increase in the disagreement of the pool of workers? Recall that optimistic team members will work harder early on if they have pessimistic coworkers, while the presence of optimists makes pessimists less productive initially. If negative sorting is to be preferred to positive sorting, it must be that the first force prevails. This happens because optimists expect any change in coworkers' future effort to be more payoff-relevant than pessimists, as, by definition, they expect higher returns on effort than pessimists. As a result, optimists face stronger persuasion incentives, and their extra motivation more than compensates for the negative effort adjustment of pessimists. The intuition for why disagreement in the population of workers increases the aggregate value of two-member team production is that when the population distribution is symmetric and full support, more disagreement in the workforce results in more within-team disagreement in the optimal transport solution, while teams' optimism remains unaffected. 

Finally, we build on the intuition of the single-technology case to study the implications of disagreement when two production technologies are available. When both technologies are similarly productive, a two-member team disagreeing over which technology works best is more productive than any team of like-minded coworkers, including one that is maximally optimistic about both technologies. To grasp an intuition of why this \textit{competition of ideas} can be particularly useful, note that disagreement over the best technology motivates each coworker to invest extra effort in their preferred technology early on in the game, to obtain the early successes that would persuade their coworker to abandon their suboptimal technology and join efforts. This is in contrast with the single technology case, where disagreement always motivates one player at the cost of making another slack off, and where the most productive team is maximally optimistic. Finally, we show that the same kind of competition of ideas can also alleviate the externality problem arising if players hold correct beliefs. Disagreement pushes effort closer to the first-best level and increases the expected welfare of each player.

All in all, our analysis suggests that disagreement could be beneficial if economic agents have an incentive to persuade each other, and their persuasion technology is productive. As in the iPhone and Metallica examples, the disagreement dividend lies in parties' determination to spread their own view and bring others on board.

\subsection{Literature Review}\label{Lit} 

At a high level, our paper is related to the economic literature on team diversity \citep[e.g.,][]{Hong2001,Hong2004,Page,Dong2022}. We highlight how the interaction between coworkers' strategic incentives and their perspectives on the production process can define whether team members with different views will, on average, outperform homogeneous teams. In contrast with a large part of the diversity literature, the driving force of our results is not just team heterogeneity, but players' awareness about disagreement: the benefits from disagreement highlighted in this paper vanish if each player behaved under the (wrong) assumption that others shared their prior belief. Hence, we highlight the importance of open disagreement. 

Some of our ideas are related to the literature on Bayesian learning with different priors \citep[e.g.,][]{Hart2020,Kartik2021}, and its applications to principal-agent problems of evidence collection \citep[e.g.,][]{Che2009,VanDerSteen} and delegation \citep[e.g.,][]{Hirsch2016}. In particular, \citet{Che2009} and \citet{VanDerSteen} show that disagreement with the principal can motivate an expert with aligned preferences to collect more evidence.\footnote{\citet{VanDerSteen} also finds that disagreement between a manager and an employee tends to reduce delegation, hinder motivation, and lower satisfaction. By focusing on different strategic incentives, we provide a complementary, more optimistic picture of the effects of disagreement on team effort and production.} We complement this literature with an analysis of team formation and teamwork, where agents persuade each other through successful production histories. Importantly, we show why an output-maximizing manager could benefit from forming disagreeing teams even when she is perfectly informed about the state and therefore does not value additional information. Additionally, our analysis suggests that the persuasion incentives of disagreeing coworkers might reduce the average output of skeptical team members.

Our analysis shares qualitative similarities with the literature on exponential bandits \citep[e.g.,][]{Keller2005,Moroni21}, and in particular \citet{Dong2018}. By assuming the unobservability of effort, we shut down signaling,  the driving mechanism of \citet{Dong2018}, isolate the effect of persuasion through output, absent in their paper, and reach different conclusions.\footnote{\citet{Dong2018} finds that asymmetric information about the risky arm can increase experimentation in a team. Her results are driven by informational asymmetries and signaling instead of open disagreement and persuasion. In \citet{Dong2018}, the player with bad news increases effort in order to pool with the player with good news. In our model, the optimistic player takes the lead, working harder to bring the skeptic on board.} 

The paper is organized as follows. In section \ref{model}, we outline the model setup. In sections \ref{onetech} and \ref{matching}, we analyze the single production technology case, and in section \ref{twotech}, we extend the analysis to the multiple technology case. Finally, in section \ref{exit}, we draw the conclusions.

\section{Model} \label{model}
This section describes the production game. First, we introduce the game elements and the equilibrium concept used to solve for the solution. We then describe the metrics of team optimism, disagreement, and performance used to compare the characteristics and outcomes of different teams. 

\subsection{The Production Game} \label{game}

A team of coworkers engages in productive activity over two periods. In each period \(t=1,2\), each team member \(i = 1, ..., N\) chooses how much effort \(e^i_t\in[0,1]\) to allocate to production. The cost of effort is quadratic\footnote{The functional form assumption comes with the benefit of nice closed-form solutions for equilibrium effort levels, and allows us to better isolate the productivity distortions due to disagreement. Our main results generalize to convex cost functions with weakly negative third derivative.} and described by \(c(e^i_t) = \frac{1}{2}(e_t^i)^2\) and, for any level of aggregate effort \(e_t = \sum_{i = 1}^Ne^i_t\) exerted by the team in period \(t\), the productive activity yields a return \(\omega\in \{R,r\}\) with probability \(\min\{e_t,1\}\) and no return with the remaining probability.\footnote{The assumption that the probability of a return increases linearly in the aggregate team effort shuts down effort complementarities due to the production technology, and allows us to isolate the information-driven implications of disagreement that are the focus of this paper. See \citet{Prat2002} for a study of how team heterogeneity affects productivity in the presence of technology-induced effort complementarities.} We denote by \(y_t\in\{R,r,0\}\) the value of team production in period \(t\), where \(y_t = 0\) means that the team effort was not successful. We maintain the assumption that, conditional on effort levels, the arrival of returns is independent across periods. 

The type of return \(\omega\) of the production technology is fixed throughout the game and initially uncertain. When effort is successful, high-return technologies yield \(\omega = R\), while low-return technologies yield \(\omega = r\), with \(R>r>0\). Player \(i\) attaches prior probability \(p^i\in [0,1]\) to \(\omega = R\), and players can start the game with different priors. Dispensing with the common prior hypothesis allows us to study how the game outcomes depend on initial team disagreement. Furthermore, we assume that players know each other's priors: whenever there is disagreement in the team, it is ``open disagreement.'' Note that this awareness of prior disagreement is not inconsistent with \citet{Aumann}'s impossibility of ``agreeing to disagree,'' as such a result holds for posterior beliefs when players hold a common prior.

The timing of the game is as follows. At the beginning of each \(t=1,2\), players simultaneously choose their effort levels (unobserved to their coworkers) and incur their personal effort cost. At the end of the period, each team member \(i = 1,..., N\) receives a payoff equal to \(u(y_t,e^i_t) = \alpha y_t - \frac{1}{2}(e^i_t)^2\), where \(\alpha\in (0,1]\) captures the rival vs public nature of the good produced. Whenever the team receives a return, so that \(y = \omega\), each team member learns \(\omega\) and updates her beliefs accordingly. More specifically, we make the following assumption about belief updating at the end of the first period.

\begin{assumption}[Belief Updating]
Let \(\pi^i\) denote the probability that \(\omega = R\) according to the belief of player \(i = 1,..., N\) at the beginning of \(t=2\). If \(y_1 >0\) then \(\pi^i = \mathbbm{1}(y_1 = R)\). If \(y_1 = 0\) then \(\pi^i = p^i\), where \(p^i\) is the probability that \(\omega = R\) according to the prior of player \(i\).
\end{assumption}

The assumption requires that, after observing a high (low) return, each player assigns probability \(1\) to the technology being of the high-return (low-return) type and that, in the absence of a return, players do not update their prior. This requirement allows for Bayesian players, as well as players following other updating rules, including paradigm-shift rules \citep{Ortoleva2012, Galperti2019}.

\paragraph{Equilibrium Concept} We look for a Markov Perfect equilibrium of the game, such that players' strategies are allowed to depend on the period \(t\) and on the beliefs held by each player at the beginning of that period. In particular, let a strategy for player \(i\) consist of an effort rule pair \(s^i = (s^i_1,s^i_2)\), where for \(t = 1,2\), effort rule \(s^i_t:[0,1]^N\to [0,\infty)\) specifies the effort level exerted by player \(i\) in period \(t\) as a function of players' beliefs at the beginning of \(t\). A strategy profile \(s = (s^i)_{i = 1}^N\) is an equilibrium of the game if and only if the following conditions are met for each \(i = 1,..., N\):

\begin{enumerate}[(i)]
\item \(s^i_2\) maximizes \(i\)'s expected second-period payoff given \(s^{-i}_2\). Formally, let \(P(e) = \min\{1,e\}\). For each posterior belief profile \(\pi = (\pi^1,...,\pi^N) \in [0,1]^N\), it has
\begin{equation*}
s^i_2(\pi) \in \arg\max_{e^i\ge 0} \left\{\alpha\mathbb{E}_{\pi^i}[\omega]P\left(e^i + \sum_{j\ne i}s^j_2(\pi)\right) - c(e^i)\right\},
\end{equation*}
where \(\mathbb{E}_{\pi^i}\) denotes the expectation operator based on belief \(\pi^i\).
\item \(s^i_1\) maximizes the present value of the sum of \(i\)'s present and future payoffs, given \(s^{-i}_1\) and \(s_2\). Formally, let \(V^i_{s_2}(\pi) = \alpha\mathbb{E}_{\pi^i_2}[\omega]P\left(\sum_{j = 1}^N s^j_2(\pi)\right) - c(s^i_2(\pi))\). For each prior belief profile \(p = (p^1,...,p^N) \in [0,1]^N\), it has
\begin{equation*}
s^i_1(p) \in \arg\max_{e^i\ge 0} \left\{\alpha\mathbb{E}_{p^i}[\omega]P\left(e^i + \sum_{j\ne i}s^j_1(p)\right) - c(e^i) + \beta\mathbb{E}_{p^i}\left[V^i_{s_2}(\pi)\Bigg|e^i + \sum_{j\ne i}s^j_1(p)\right]\right\},
\end{equation*}
where \(\beta\in(0,1]\) is a payoff discount factor.
\end{enumerate}

A few clarifications are due. First, note that we are ignoring equilibria in mixed strategies, as well as effort rules that are contingent on past actions. Both assumptions are without loss of generality: mixed strategies are never used in equilibrium, and, in the second period, conditioning effort levels on past actions is a strictly dominated strategy. Second, expressing first-period strategies as functions of prior beliefs might look unnecessary, as priors are typically fixed and known at the beginning of the game. In the context of this paper, this formulation comes in handy given that our analysis studies how team equilibrium performance depends on the distribution of team members' initial beliefs.

When the aggregate effort \(e_t\) reaches 1, any additional effort does not increase the probability of obtaining a return. To rule out that this boundary case arises in equilibrium, we make the following assumption.

\begin{assumption}\label{assbound}
The aggregate value of returns is not too large, \(2NR < 1.\)
\end{assumption}
As we shall see, the assumption is sufficient (but not necessary) to guarantee that, in equilibrium, the team effort fails with positive probability. It does not otherwise influence our results. Before proceeding with the equilibrium analysis, the next section introduces our metrics of team optimism, team disagreement, and team performance. 

\subsection{Team Metrics}

Throughout the paper, we will compare the equilibrium performance of teams that differ only in one characteristic, the initial beliefs of their members (henceforth, their \textit{type}). We interpret \(p^i\) as a measure of (initial) optimism of team member \(i\), as a higher \(p^i\) means that she attaches a higher probability to the technology being of the high-return type. For the same reason, we say that a team member \(i\) is more \textit{optimistic} than team member \(j\) if \(p^i>p^j\) and that team member \(i\) is more \textit{pessimistic} than team member \(j\) if \(p^i<p^j\). Finally, if \(p^i = p^j\) then the two team members are \textit{like-minded}, while they \textit{disagree} if \(p^i \ne p^j\), so that \((p^i-p^j)^2\) can be used as a measure of disagreement between the two team members.

We use these intuitions to define metrics of team optimism and team disagreement. For each \(p,p^{\prime}\in [0,1]^N\), we measure the optimism \(\mathcal{O}(p)\) of a team if type \(p\) by the average optimism of its members.
\begin{equation*}
\mathcal{O}(p) = \bar{p}
\end{equation*}
for \(\bar{p} = \frac{1}{N}\sum_{i=1}^Np^i\), so that if \(\mathcal{O}(p)>\mathcal{O}(p^\prime)\) a team of type \(p\) is more optimistic than a team of type \(p^{\prime}\). We measure team disagreement \(\mathcal{D}(p)\) as the average disagreement between any two team members,
\begin{equation*}
\mathcal{D}(p) = \frac{2}{N(N-1)}\sum_{i=1}^{N-1}\sum_{j=i+1}^N(p^i-p^j)^2.
\end{equation*}
If \(\mathcal{D}(p)>\mathcal{D}(p^\prime)\) we say that team \(p\) disagrees more than team \(p^\prime\), and if \(D(p) = 0\), then we say that \(p\) is a like-minded team.

Finally, we measure the team performance of \(p\) at the unique\footnote{As we show in the next section, the game admits a unique (pure strategy) equilibrium.} equilibrium profile \(s\) as the expected aggregate output \(\mathcal{Y}_{p^\star}(p)\) of the team when its members follow the equilibrium strategies and the expectation is evaluated at the prior \(p^\star\in[0,1]\) of an analyst or team manager, 
\begin{equation*}
\mathcal{Y}_{p^\star}(p) = \mathbb{E}_{p^\star}\left[\omega\sum_{i}^N\left(s_1^i(p) + \beta s_2^i(\pi)\right)\right].
\end{equation*}
If \(\mathcal{Y}_{p^\star}(p)>\mathcal{Y}_{p^\star}(p^\prime)\), team \(p\) is more productive than team \(p^{\prime}\).

In the next sections, we use these metrics to relate team optimism, disagreement, and performance.
\section{The Disagreement Dividend} \label{onetech}

How does disagreement affect coworkers' effort? Does team disagreement influence team productivity? If yes, is it desirable? This section addresses these questions through the lens of our parsimonious model. 

Before proceeding, it is useful to define some auxiliary variables. Let \(\tilde{R} = \alpha R\) and \(\tilde{r} = \alpha r\) be the returns accruing to any team member when the team obtains returns \(R\) and \(r\) respectively. Similarly, let \(\tilde{\omega} = \alpha\omega\). Finally, let \(\Delta = \tilde{R} - \tilde{r}\). 

Note that at \(t = 2\) team member \(i = 1,..., N\) with posterior \(\pi^i\) gains an expected payoff equal to \(\mathbb{E}_{\pi^i}[\tilde{\omega}]P(e^i_2) - \frac{1}{2}e^i_2\). If \(e^{-i}_2 \le 1-\tilde{R}\), her payoff is maximized at \(e^\star_2(\pi^i) = \mathbb{E}_{\pi^i}[\tilde{\omega}]\), which is at most \(\tilde{R}\). If instead \(e^{-i}_2 > 1-\tilde{R}\), she chooses an effort level strictly less than \(\tilde{R}\). By assumption \ref{assbound}, it has \(N\tilde{R}<1\). Hence, it must be that \(e^{-i}_2 \le 1-\tilde{R}\). This means that in any equilibrium, the second period effort rule of player \(i = 1, ..., N\) is 
\begin{equation}
    s^i_2(\pi) = e^\star_2(\pi^i) = \mathbb{E}_{\pi^i}[\tilde{\omega}].
\end{equation}
Since \(\mathbb{E}_{\pi^i}[\tilde{\omega}]=\pi^i\Delta + \tilde{r}\), the above expression tells us that in the second period, the effort of a team member is strictly increasing in her posterior belief \(\pi^i\): more optimistic players will work harder in the second period than pessimistic ones. But hard work increases the probability that the team will succeed in the second period (\(y_2>0\)), so each member of the team would want her colleagues to be as optimistic as possible at \(t = 2\). Our first result, captured by the following proposition, builds on this intuition. 

\begin{proposition}\label{propeffort}
The first-period effort of a team member is decreasing in the optimism of her team. Formally, let \(p,p^{\prime}\in [0,1]^N\) such that \(p^i = p^{\prime i}\). It holds,
\begin{equation}
s^i_1(p) > s^i_1(p^\prime) \iff \mathcal{O}(p) < \mathcal{O}(p^\prime).
\end{equation}
As a result, at \(t=1\), a team member exerts more (less) first-period effort in a team that is less (more) optimistic than her than in a like-minded team.
\end{proposition}

We illustrate the intuition of the proposition by focusing on a team of two coworkers, Ann and Bob (\(i=1\) and \(i=2\), respectively). After some manipulations, the first-order condition on the maximization problem faced by \(i=1,2\) in the first period yields the following decomposition,

\begin{equation}\label{effortdecomposition}
\underbrace{e^i_1(p^1,p^2)}_{\substack{\text{marginal cost} \\ \text{of effort}}} = \underbrace{\mathbb{E}_{p^i}[\tilde{\omega}]}_{\substack{\text{expected current} \\ \text{gain from} \\ \text{return arrival}}} + \underbrace{\beta\mathbb{E}_{p^i}[\tilde{\omega}](e^\star_2(p^i)-e^\star_2(p^j))}_{\substack{\text{expected future gain from bringing} \\ \text{coworker on the same page}}} + \underbrace{\frac{3\beta}{2}\textit{Var}_{p^i}(\tilde{\omega})}_{\substack{\text{expected future gain} \\ \text{from uncertainty} \\ \text{reduction}}},
\end{equation}
where the left-hand side is the marginal cost of effort and the right-hand side is the marginal benefit from effort.\footnote{Also in this case, assumption \ref{assbound} ensures that \(e^i_1(p^1,p^2) < 1\).} Note that, because the probability of a success in \(t=1\) increases linearly with \(e^i_1\), the marginal benefit from effort is equal to \(i\)'s expected payoff gain from a team success (\(y_1 > 0\)) instead of a team failure (\(y_1 = 0\)) at \(t=1\). 

A first-period success comes with two types of gains. The first is a return arrival in the present, with expected value equal to the first term on the right-hand side of equation \ref{effortdecomposition}. The second benefit is the expected future payoff gain from learning.

From the perspective of player \(i\), learning has two implications: (i) initial disagreement is resolved, and (ii) the prior belief \(p^i\) is updated, leading to a better allocation of effort. Equation  \ref{effortdecomposition} decomposes the effect of learning in these two components. Resolving initial disagreement means that \(j's\) belief is corrected from \(p^j\) to \(p^i\), yielding a change in her future effort from \(e^\star_2(p^j)\) to \(e^\star_2(p^i)\). With this belief correction, the present value of \(i\)'s expected second-period payoff changes by the second term on the right-hand side of equation \ref{effortdecomposition}. Note that \(e^\star_2(p^i)-e^\star_2(p^j)\) is \(i\)'s expectation about  \(j\)'s effort change from learning: since \(\mathbb{E}_{p^i}\left[\pi^j|y_1>0\right] = p^i\), it has
\begin{align*} 
e^\star_2\left(p^i\right)-e^{\star}_2\left(p^j\right) &= \Delta\left(p^i-p^j\right) \\
&= \Delta\underbrace{\left(\mathbb{E}_{p^i}\left[\pi^j|y_1>0\right]-p^j\right)}_{\substack{\text{i's expectation of j's belief} \\ \text{change from learning}}} \\
& = \underbrace{\mathbb{E}_{p^i}\left[e^\star_2\left(\pi^j\right)-e^{\star}_2\left(p^j\right)|y_1>0\right]}_{\substack{\text{i's expectation of j's effort} \\ \text{change from learning}}}
\end{align*}
This highlights the key persuasion mechanism in place in a disagreeing team. When Ann is more optimistic than Bob, she expects a return arrival, on average, to confirm her belief and make Bob more optimistic. Such an increase in optimism motivates Bob to work harder in the future, and Ann benefits from Bob's future hard work. Hence, Bob's pessimism motivates Ann to work harder early on to increase the likelihood of a return in the first period, compared to the case when Bob is like-minded. The reverse implication works for Bob: as he expects a return to make Ann more pessimistic, he has the incentive to work less in the first period than when paired with a like-minded coworker. 

Finally, learning \(\tilde{\omega}\) comes with the benefit of updating \(p^i\), an effect captured by the third term on the right-hand side of equation \ref{effortdecomposition}. While \(i\) expects both team members to exert effort \(e^\star_2(p^i)\) after a return arrival at \(t=1\), she also anticipates that learning \(\tilde{\omega}\) will improve the future allocation of effort across states: the team will work harder when the technology yields high returns than when it yields low returns.  This benefit from learning is proportional to \(\textit{Var}_{p^i}(\tilde{\omega})\), \(i\)'s prior uncertainty about the technology type. 

We have found that, when Ann is more optimistic than Bob, then (i) Ann works harder early on when paired with Bob than with a like-minded coworker; and (ii) Bob works harder early on when paired with a like-minded coworker than with Ann. Before presenting our next result on the comparison between disagreeing teams, it is worth pointing out that the upward first-period distortion (i) more than compensates the downward distortion (ii). Let \(p^1>p^2\). From equation \ref{effortdecomposition}, we have that 
\begin{equation*}
\sum_{i=1,2} e^i_1(p^1,p^2) - \sum_{i=1,2} e^i_1(p^i,p^i) = \beta(\mathbb{E}_{p^1}[\tilde{\omega}]-\mathbb{E}_{p^2}[\tilde{\omega}])(e^\star_2(p^1)-e^\star_2(p^2))>0.
\end{equation*}
Intuitively, Ann's incentive to distort first-period effort is stronger than Bob's, because convincing her coworker is worth more for Ann, as she's more optimistic about the marginal returns on effort, \(\mathbb{E}_{p^1}[\tilde{\omega}] > \mathbb{E}_{p^2}[\tilde{\omega}]\). This observation, and the following result on team performance lie the foundations for the optimal matching analysis presented in the next section.

\begin{proposition}\label{propdis}
Holding team optimism constant, team performance is increasing in team disagreement. More specifically, let teams \(p,p^\prime\in [0,1]^N\) be such that \(\mathcal{O}(p) = \mathcal{O}(p^\prime)\) and \(\mathcal{D}(p^\prime) = 0\), so that team \(p^\prime\) is like-minded. For each team manager's prior \(p^\star\in[0,1]\) it has
\begin{equation*}
\mathcal{Y}_{p^\star}(p) = \mathcal{Y}_{p^\star}(p^\prime) + \underbrace{\kappa^\mathcal{O}_{p^\star}\mathcal{D}(p)}_{\substack{\text{disagreement} \\ \text{dividend}}}
\end{equation*}
where \(\kappa_{p^\star}^{\mathcal{O}}\) is strictly positive and only depends on \(\mathcal{O}(p)\) and \(p^\star\).
\end{proposition}
The explicit expression of \(\kappa_{p^\star}^{\mathcal{O}}\) is provided in the appendix. Holding constant team optimism, team disagreement boosts production, giving rise to a \textit{disagreement dividend}. This has two main implications. First, when we compare two equally optimistic teams, the one that exhibits more disagreement is expected to be more productive. Second, while it can be shown that for a fixed level of disagreement team performance is always increasing in team optimism, an implication of proposition \ref{propdis} is that disagreement can be a substitute for optimism: under some conditions, a disagreeing team outperforms teams formed of more optimistic but more like-minded agents. 

The disagreement dividend originates from the persuasion incentives captured by the second term on the right-hand side of equation \ref{effortdecomposition}. When the team is composed of \(N\) members, the corresponding term for player \(i\) becomes
\(\beta\mathbb{E}_{p^i}[\tilde{\omega}]\sum_{j\ne i}(e^\star_2(p^i)-e^\star_2(p^j))\). Adding across team members,
\begin{align*}
\beta\sum_{i=1}^N\mathbb{E}_{p^i}[\tilde{\omega}]\sum_{j\ne i}(e^\star_2(p^i)-e^\star_2(p^j)) & = \beta\sum_{i=1}^N\sum_{j\ne i}e^\star_2(p^i)(e^\star_2(p^i)-e^\star_2(p^j)) \\
&= \frac{\beta}{2}\sum_{i=1}^N\sum_{j\ne i}(e^\star_2(p^i)-e^\star_2(p^j))^2 \\
&= \frac{\beta\Delta^2}{2}\sum_{i=1}^N\sum_{j\ne i}(p^i-p^j)^2 \\
&= \beta\Delta^2\frac{N(N-1)}{2}\mathcal{D}(p).
\end{align*}
Hence, the larger the team disagreement, the stronger, on average, the motivation to persuade coworkers through hard work in the first period. This effort boost, as we show in the appendix, drives up the aggregate expected output of disagreeing teams. 

Finally, note that awareness of disagreement plays a crucial role in generating the disagreement dividend of proposition \ref{propdis}. Intuitively, only team members who know about the different beliefs of their coworkers can factor into their optimal effort choice the benefit of changing others' beliefs. The fundamental role of disagreement awareness stands out in our setup: if each member \(i\) of a disagreeing team \(p\) with fixed optimism \(\mathcal{O}(p)\) chose initial effort levels as if \(p^j=p^i\) for each \(j\ne i\), then the team's expected output would be \textit{decreasing} --- not increasing --- in disagreement \(\mathcal{D}(p)\).

\section{Team Formation}\label{matching}

We have shown that disagreement has opposite motivational effects for optimistic and pessimistic coworkers, and that a disagreeing team performs better than a more like-minded and similarly optimistic one. We now build on these two observations to study an optimal team formation problem faced by a manager who wants to allocate coworkers into teams, intending to maximize the aggregate production of her workforce.

The workforce consists of a continuum of employees with mass normalized to one. The manager, \(M\), needs to allocate the employees in a continuum of two-member teams. The two members of a team will simultaneously engage in the two-period productive activity described by the game studied in the previous section, for \(N = 2\). It is common knowledge that team technology types \(\omega\) are independent and identically distributed across teams, and employees only differ by their commonly known prior level of optimism. 
In particular, the workforce's prior beliefs that a technology is of type \(R\) follow a distribution with continuous cumulative distribution function \(F\) with support \([0,1]\), and that the \(M\)'s prior belief is \(p^\star\in [0,1]\). Finally, conditional on effort levels, return arrivals are independent across teams and periods.

The matching task faced by \(M\) can be formalized as solving the following optimal transport problem,
\begin{equation}\label{maxtransport}
\max_{\gamma\in\Gamma(F,F)}\frac{1}{2}\int_{[0,1]^2}Y_{p^\star}(p,p^\prime)d\gamma(p,p^\prime),
\end{equation}
where \(\Gamma(F,F)\) is the set of joint probability measures with both marginals equal to \(F\).

How should the manager pair coworkers? Should they create like-minded teams and rely on the productive performance of the most optimistic teams? Will the manager aim at maintaining an intermediate level of disagreement in all teams, or should they form some teams with high disagreement and some with high levels of like-mindedness?

The next proposition suggests that the optimal matching is negative assortative, with the most optimistic employees paired together with the most pessimistic ones, and moderately optimistic ones paired together with moderately pessimistic ones. 

\begin{proposition}\label{optimaltransport}
The manager problem has a unique solution, where employees are matched in a negative assortative way. In particular, let \(T_{F}:[0,1]\to [0,1]\) be the optimal team formation plan associated with the solution of problem \ref{maxtransport}, so that the teammate of an employee of type \(p\) has type \(T_{F}(p)\). It has
\begin{equation*}
T_{F}(p) = F^{-1}(1-F(p))
\end{equation*}
for each \(p\in[0,1]\).
\end{proposition}

In rough terms, the proposition tells us that maximally optimistic employees should be matched with maximally pessimistic ones, and, in general an employee who is more optimistic than exactly \(q\) percent of the workforce should be matched with one who is more pessimistic than exactly \(q\) percent of the workforce.

What gives rise to this negative sorting? In the appendix we show that 

\begin{equation*}
\frac{\partial^2 e^i_1(p^1,p^2)}{\partial p^1\partial p^2}<0 \text{ for \(i = 1,2\)} \implies \frac{\partial \mathcal{Y}_{p^\star}(p^1,p^2)}{\partial p^1\partial p^2} < 0,
\end{equation*}
so that the expected team production is strictly submodular in team members' initial beliefs. This (strict) submodularity of expected output gives rise to the negative sorting result. By pairing Ann and Bob together, we make sure that the player who expects her coworker to increase effort the most in the event of an early success is also the one who values an increase in effort the most (i.e., Ann). Symmetrically, we also guarantee that the player who expects her coworker to reduce effort the most in the event of an early success is also the one who values such a change in effort the least (i.e., Bob). The virtuous effects on first-period effort translate to overall team performance, implying the optimality of negative sorting.

Next, we ask how disagreement in the full workforce influences the value of problem \ref{maxtransport}, that is, the overall expected output of the firm. To draw an analogy with our measure of team disagreement, we measure workforce disagreement by 
\begin{equation*}
\bar{\mathcal{D}}(F) = \mathbb{E}_{F}[(p - p^\prime)^2],    
\end{equation*}
for \(p\) and \(p^\prime\) independently drawn from \(F\). In other words, workforce disagreement refers to the expected disagreement between two randomly selected employees. 

Let \(\mathcal{V}_{p^\star} (F)= \frac{1}{2}\int_{[0,1]}\mathcal{Y}_{p^\star}(p,T_F(p))dF(p)\) be the value of problem \ref{maxtransport}, henceforth the \textit{value of workforce} \(F\). The next proposition tells us that, under symmetry conditions, the aggregate productivity of the optimal teams is increasing in workforce disagreement.  

\begin{proposition}\label{workforcedis}
When workforce optimism is distributed symmetrically, the value of the workforce increases in its disagreement. Formally, let \(F,G\) be two symmetric continuous distributions with support \([0,1]\). It has
\begin{equation*}
\mathcal{V}_{p^\star}(F)>\mathcal{V}_{p^\star}(G) \iff \bar{\mathcal{D}}(F) > \bar{\mathcal{D}}(G).
\end{equation*}
\end{proposition}

The result follows from proposition \ref{propdis}. In fact, note that 
\begin{align*}
\mathcal{V}_{p^\star}(F) - \mathcal{V}_{p^\star}(G) &= \frac{1}{2}\left(\int_{[0,1]}\mathcal{Y}_{p^\star}(p,T_F(p))dF(p) - \int_{[0,1]}\mathcal{Y}_{p^\star}(p,T_G(p))dG(p)\right) \\
&= \frac{k^{\mathcal{O}}_{p^\star}}{2}\left(\int_{[0,1]}\mathcal{D}(p,T_F(p))dF(p) - \int_{[0,1]}\mathcal{D}(p,T_G(p))dG(p)\right) \\
&= k^{\mathcal{O}}_{p^\star}\left(\bar{\mathcal{D}}(F) - \bar{\mathcal{D}}(G)\right),
\end{align*}
where the second inequality holds because symmetry and full support imply that all teams in the optimal match have the same optimism (both within and between workforces), and the third holds because, when all teams in the workforce have the same level of optimism, average team disagreement can be shown to equal workforce disagreement.

So far, we have shown that disagreement motivates the effort of optimists while discouraging pessimists. We have argued that it can serve as a substitute for team optimism as an input for team production, giving rise to a disagreement dividend. Finally, we illustrated that a manager responsible for assigning a large pool of workers to two-member teams should pair teammates in a negative assortative way, and that the aggregate product of teams increases in workforce disagreement. In the next section, we extend the analysis to allow for multiple production technologies. We show that, under certain conditions, the most productive team disagrees about which technology works best. Besides boosting team production, this disagreement makes all coworkers better off.

\section{A Tale of Two Methods} \label{twotech}

We now study the variation of this game where, in each period, each team member can choose to operate one of two different production technologies, \(A\) and \(B\). For simplicity, we focus on the case with \(N=2\) team members.\footnote{The main results for the two-technology case are robust to allowing for \(N>2\) and \(K > 2\) production technologies.} Technologies are characterized by types \(\omega_A,\omega_B\in\{R,r\}\) with independent realizations, so that player \(i\) opinion is now characterized by vector \(p^i = (p^i_A,p^i_B)\in [0,1]^2\). If, in period \(t\), both workers operate the same technology, \(k\in \{A,B\}\), then the team output is \(y_t = \omega_k\) with probability \(P(e_t^1+e_t^2)\), and \(y_k = 0\) with the remaining probability. If they operate different technologies --- e.g., team member \(1\) chooses \(A\) and team member \(2\) chooses \(B\) --- then \(y_t=\omega_A\) with probability \(P(e^1_t)(1-P(e^2_t))\), \(y_t = \omega_B\) with probability \(P(e^2_t)(1-P(e^1_t))\), \(y_t = \omega_A + \omega_B\) with probability \(P(e^1_t)P(e^2_t)\), and \(y_t = 0\) with the remaining probability. Conditional on effort levels, return arrivals are independent across technologies and periods, and we assume that the period \(t\) utility from team output \(y_t\) and the costs of effort are as specified in section \ref{game}. For analogy with the one-technology setup, we measure team disagreement by \(D(p) = ||p^1-p^2||^2\).

The multi-technology setup admits many possible interpretations. For instance, our group could consist of board members of a company trying to decide which business project to prioritize between a number of promising, but uncertain alternatives. They could be investors deciding between two competing investment portfolios. They could be academics disagreeing over which research method, or theory, will yield the best answers to a research question. They might be collaborating artists, disagreeing over the best way to create an artwork or song. In all cases, teammates have access to alternative methods of production. Importantly, we maintain the assumption that the players' interests are aligned, to some degree: each of them benefits if the other adopts the best production method --- no matter how strong the disagreement over what such a method is.

In this two-technology scenario, the intuition of the previous analyses carries through when we focus on disagreement between generally optimistic and generally pessimistic players, e.g., characterized by beliefs \((1,1)\) and \((0,0)\) respectively. However, extending the analysis to more than two production technologies allows us to capture a second type of disagreement, whereby coworkers cannot be ranked in terms of their level of optimism, but disagree about which production method works best --- as, for instance, when their beliefs are \((1,0)\) and \((0,1)\). This type of disagreement, which we call \textit{competition of ideas}, has an intuitive appeal: when coworkers believe that only one technology is promising and disagree on which one, they are motivated to work harder early on, to obtain the breakthrough that can persuade their coworker to switch to the technology they are most optimistic about.  

Before stating our results formally, it is useful to augment our strategy and equilibrium definition to account for technology choices. In the multi-technology context, a strategy \(s^i\) for player \(i = 1,2\) is pair of time and belief contingent plans \((s^i_1,s^i_2)\) such that, for each period \(t = 1,2\), \(s^t_i: [0,1]^4\to \{A,B\}\times[0,1]\) specifies a technology choice and effort level for each belief pair held by the two team members at the beginning of the period. The equilibrium concept is the same as that utilized in the previous sections, except that now, in each period, players maximize over two variables, the technology operated and the effort exerted in that technology. As the game with two technologies admits multiple equilibria \(s\), we define team performance as the expected output in the most productive equilibrium for a team of type \(p\in[0,1]^4\). Formally, if \(S\) is the set of equilibria, and \(\mathcal{Y}^s_{p^\star}: [0,1]^4\to \mathbb{R}\) is the team performance function at equilibrium \(s\in S\), \(\mathcal{Y}_{p^\star}(p) = \max_{s\in S} \mathcal{Y}^{s}_{p^\star}(p)\). Finally, let \(C\) define the set of beliefs with maximal competition of ideas, \(C = \{((1,0),(0,1)), ((0,1), (1,0))\}\).

\begin{proposition}\label{hdis}
Let \(p^\star = (p,p)\), for \(p\in[0,1]\). If \(\beta\) is not too large or \(p\) is large enough, a team with competing ideas performs better than any like-minded team. Formally,

\begin{enumerate}[(i)]
\item If \(\beta<\frac{\tilde{r}}{\Delta}\), then 
\begin{equation*}
\mathcal{Y}_{p^\star}(p^C)>\mathcal{Y}_{p^\star}(p^L) \quad\forall p\in[0,1];
\end{equation*}
\item If \(\beta\ge\frac{\tilde{r}}{\Delta}\), then there exists a threshold \(\bar{p}\in[0,1)\) such that, for all \((p^C, p^L)\in C\times L\),
\begin{align*}
\mathcal{Y}_{p^\star}(p^C)>\mathcal{Y}_{p^\star}(p^L) \iff p>\bar{p},
\end{align*}
with \(\mathcal{Y}_{p^\star}(p^C) = \mathcal{Y}_{p^\star}(p^L)\) if and only if \(p=\bar{p}\) and \(\bar{p} = 0\) if and only if \(\beta = \frac{\tilde{r}}{\Delta}\).
\end{enumerate}
\end{proposition}
The intuition of the proposition is the following. In stark contrast with the result of proposition \ref{propeffort}, when there is competition of ideas, pessimism about the technology favored by the coworker does not discourage effort, but motivates a team member to work harder. In particular, we show that at \(t=1\), each worker always operates the technology she's optimistic about, anticipating that an early success will persuade her teammate to switch to her ``superior'' production approach in the future. More formally, the first order condition for the first-period equilibrium effort level of \(i=1,2\) becomes

\begin{equation}\label{switch}
\underbrace{e^i_1(p^C)}_{\substack{\text{marginal cost} \\ \text{of effort}}} = \underbrace{\tilde{R}}_{\substack{\text{expected current} \\ \text{gain from} \\ \text{return arrival}}} + \underbrace{\beta\tilde{R}\frac{(1+\tilde{r})\Delta}{1-\beta\tilde{r}\Delta}}_{\substack{\text{expected future} \\ \text{gain from coworker's} \\ \text{technology switch}}}.
\end{equation}
The second term on the right-hand side of equation \ref{switch} captures \(i\)'s expectation of \(j\)'s output increase when \(j\) reallocates her second-period effort from the technology \(i\) is pessimistic about to the one she is optimistic about. If, according to \(p^\star\), both production technologies are equally productive, an increase in both players' first-period effort translates into higher expected total team output, regardless of the technology that is used. The result of part (ii) for \(\beta>\frac{\tilde{r}}{\Delta}\) is unchanged if we compare the expected output of teams \(p^C\) and \(p^L\) in any equilibria, as opposed to most productive ones. Additionally, this same intuition generalizes when technologies are not identical, but sufficiently similar.

So far, we have shown that certain types of disagreement have positive effects on team output. Our last finding suggests that the same disagreement can increse both team members' expected utility. Denote by \(\mathcal{W}^i_{p^\star}(p)\) the payoff expected to accrue to member \(i\) of a team with initial beliefs \(p\in [0,1]^4\), evaluated according to belief \(p^\star\) at a most productive equilibrium \(\hat{s}\in\arg\max_{s\in S} \mathcal{Y}^{s}_{p^\star}(p)\).

\begin{proposition}\label{welfare}
Let \(p^\star = (1,1)\). The competition of ideas gives a greater expected payoff for all team members than like-mindedness. Formally, 
\begin{equation*}
\mathcal{W}^i_{p^\star}(p^C)>\mathcal{W}^i_{p^\star}(p^L)
\end{equation*}
for all \((p^C,p^L)\in C\times L\) and \(i = 1,2\).
\end{proposition}

Why does the competition of ideas make players better off? Because, even when holding correct beliefs \(p^i = p^\star = (1,1)\), each worker's period effort choice ignores the positive externality that a return arrival has on their teammate. When the correct prior is \(p^\star = (1,1)\), players' expected period payoff is maximized if each of them allocates effort \(2\tilde{R}\) to opposite technologies. Competition of ideas does not fix the externality problem in the second period, but it reduces it in the first-period. In fact, while \(e^i_1(p^C)\) is still below \(2\tilde{R}\), it is strictly greater than the first-period effort observed in like-minded teams. 

The takeaway of this section is simple and yet, we believe, important and non-trivial. When two similarly good production technologies are available, disagreement on the best way to produce might increase group production. We focused on a mechanism that builds on the idea of different perspectives and shows that these differences can be useful even when they do not complement each other, that is, when there is \textit{competition of ideas}. When they have the common goal of increasing each other's productivity, disagreeing people will challenge each other, and work harder to prove, with their successes, that their own perspective is valid, beneficial, and worth adopting. The resulting increase in group output can make all team members better off.

\section{Conclusion} \label{exit}

We have shown that disagreement within a group of economic agents who repeatedly engage in a productive activity can increase the aggregate output of the group. This relation between disagreement and productivity helps us benchmark our results with the main findings of the theoretical literature on diversity and problem solving \citep[e.g.,][]{Hong2001,Hong2004}. In this literature, different perspectives are seen as an asset, but under two assumptions. First, team members must be able to cooperate and combine their perspectives in a productive way. Second, and related, different perspectives must not lead to different goals. 

In the paper, we have presented a mechanism that on the surface departs from the first assumption: even when different perspectives are in\textit{ conflict} with each other, leading to disagreement about the most productive approach, they might still be very useful to a team of innovators. Peers' pessimism and competition of ideas can motivate team members to work harder to prove their point. If anything, our results illustrate that some degree of ``scientific'' skepticism of each group member towards the perspective of others can be a powerful motivator: the disagreeing group should be aware that only perspectives that prove successful are eventually adopted, as this force might push them to work harder to convince others. At the same time, from a high-level point of view, our findings suggest that the benefit of disagreement should materialize if agents' ultimate goals are somewhat aligned by positive production externalities, while differences might harm if players want to reduce the productivity of others (negative externalities).

Finally, we have discussed a type of communication that --- we believe --- is sometimes overlooked by the economic literature: the type of persuasion that comes from the tangible results of economic actions, rather than from information design by a sender or the implicit informational content of a given equilibrium behavior. We believe that a deeper investigation of this force could help explain a variety of phenomena that go beyond productive incentives, including voter polarization, excessive debt accumulation, and other policy distortions.  

\bibliographystyle{apalike}
\bibliography{biblio}
\newpage
\appendix
\section{Appendix}
\vspace{.5cm}

\paragraph{Proof of proposition \ref{propeffort}}
We start by proving that in equilibrium it must be that for each \(i=1,...,N\) and \(\pi\in[0,1]^N\), team member \(i\) plays \(s^i_2(\pi) = \mathbb{E}_{\pi_i}[\tilde{\omega}] = \pi_i\Delta +\tilde{r}\). Let \(s\) be an equilibrium of the game. Fix \(i\in\{1,...,N\}\) and \(\pi\in[0,1]^N\). By our equilibrium definition, \(s^i_2(\pi)\) solves
\begin{equation*}
\max_{e^i\ge 0} \left\{\mathbb{E}_{\pi^i}[\tilde{\omega}]P\left(e^i + \sum_{j\ne i}s^j_2(\pi)\right) - \frac{1}{2}(e^i)^2\right\}.
\end{equation*}
Assume that in equilibrium \(\sum_{j\ne i}s^j_2(\pi)<1-\tilde{R}\). We can rewrite the above maximization problem as
\begin{equation}\label{maxpf1}
\max_{e^i\ge 0} \left\{\mathbb{E}_{\pi^i}[\tilde{\omega}]\left(e^i + \sum_{j\ne i}s^j_2(\pi)\right) - \frac{1}{2}(e^i)^2\right\}.
\end{equation}
Because the objective function is strictly concave, the first-order condition identifies the unique maximizer. The first order condition yields
\begin{equation}\label{eqeffort2}
s^i_2(\pi) = \mathbb{E}_{\pi^i}[\tilde{\omega}],
\end{equation}
As \(\mathbb{E}_{\pi^i}[\tilde{\omega}]\) does not depend on \(\pi^{-i}\), \(s^i_2(\pi)\) does not depend on \(\pi^{-i}\). Note that \(\mathbb{E}_{\pi_i}[\tilde{\omega}] = \pi_i\Delta +\tilde{r}\) and recall that \(\Delta = \tilde{R} -\tilde{r}\), so it holds \(s^i_2(\pi)\le \tilde{R}\). Hence, by assumption \ref{assbound}, \(\sum_{i\in I}s^i_2(\pi) < 1\), which verifies \(\sum_{j\ne i}s^j_2(\pi)<1-\tilde{R}\). Define \(e^\star_2(\pi^i) \equiv \pi_i\Delta +\tilde{r}\). The value of problem \ref{maxpf1} is 
\(V^i_{s_2}(\pi) = \mathbb{E}_{\pi^i}[\tilde{\omega}]\sum_{j = 1}^Ne^\star_2(\pi^j) - \frac{1}{2}(e^\star_2(\pi^i))^2\) and, for any prior belief \(p^i\in [0,1]\), 
\begin{align*}
\mathbb{E}_{p^i}[V^i_{s_2}(\pi)|y_1>0] &= p^i\left[\tilde{R}\sum_{j = 1}^Ne^\star_2(1) - \frac{1}{2}(e^\star_2(1))^2\right] + (1-p^i)\left[\tilde{r}\sum_{j = 1}^Ne^\star_2(0) - \frac{1}{2}(e^\star_2(0))^2\right] \\
&= \mathbb{E}_{p^i}[\tilde{\omega}]\sum_{j = 1}^N\left[p^ie^\star_2(1)+(1-p^i)e^\star_2(0)\right] + \\
&\quad + N\left[p^i(\tilde{R}-\mathbb{E}_{p^i}[\tilde{\omega}])\tilde{R} + (1-p^i)(\tilde{r}-\mathbb{E}_{p^i}[\tilde{\omega}])\tilde{r}\right] - \frac{1}{2}\mathbb{E}_{p^i}[\tilde{\omega}^2] \\
&= \mathbb{E}_{p^i}[\tilde{\omega}]\sum_{j = 1}^N\left[p^ie^\star_2(1)+(1-p^i)e^\star_2(0)\right] + NVar_{p^i}(\tilde{\omega}) - \frac{1}{2}\mathbb{E}_{p^i}[\tilde{\omega}^2] \\
&= \mathbb{E}_{p^i}[\tilde{\omega}]\sum_{j = 1}^Ne^\star_2(p^i)+ NVar_{p^i}(\tilde{\omega}) - \frac{1}{2}\mathbb{E}_{p^i}[\tilde{\omega}^2].
\end{align*}
Furthermore, it has
\begin{equation*}
\mathbb{E}_{p^i}[V^i_{s_2}(\pi)|y_1=0]= \mathbb{E}_{p^i}[\tilde{\omega}]\sum_{j = 1}^Ne^\star_2(p^j) - \frac{1}{2}(e^\star_2(p^i))^2.
\end{equation*}
Next, we want to find the (unique) equilibrium first-period effort rule \(s^i_1\) for \(i = 1,..., N\). Fix \(i \in \{1, ..., N\}\) and \(p\in [0,1]^N\). Assume that \(\sum_{j=1}^N s^j(p)\le1\). By the equilibrium definition, \(s^i_1\) must maximize the following objective function,
\begin{equation*}
\mathbb{E}_{\pi^i}[\tilde{\omega}]\left(e^i + \sum_{j\ne i}s^j_1(p)\right) - \frac{(e^i)^2}{2} + e^i\beta\mathbb{E}_{p^i}[V^i_{s_2}(\pi)|y_1>0] + (1-e^i)\beta\mathbb{E}_{p^i}[V^i_{s_2}(\pi)|y_1=0],
\end{equation*}
or, equivalently, solve
\begin{equation}\label{maxpf2}
\max_{e^i\ge 0}\left\{\mathbb{E}_{\pi^i}[\tilde{\omega}]e^i - \frac{(e^i)^2}{2} + e^i\beta\left[\mathbb{E}_{p^i}[\tilde{\omega}]\sum_{j = 1}^N(e^\star_2(p^i)-e^\star_2(p^j))+ \frac{2N-1}{2}Var_{p^i}(\tilde{\omega})\right]\right\}.
\end{equation}
Note that the objective function of problem \ref{maxpf2} is strictly concave in \(e^i\), so the first order condition identifies the unique maximizer. The first order condition leads to
\begin{equation}\label{eqeffort1}
s^i_1(p) = \mathbb{E}_{p^i}[\tilde{\omega}] + \beta\left[\mathbb{E}_{p^i}[\tilde{\omega}]\sum_{j = 1}^N(e^\star_2(p^i)-e^\star_2(p^j))+ \frac{2N-1}{2}Var_{p^i}(\tilde{\omega})\right].
\end{equation}
Note that \(s^i_1(p)\) is strictly increasing in \(p^i\) and strictly decreasing in \(p^j\) for all \(j\ne i\). When \(p^i = 1\) and \(p^j = 0\) for all \(j\ne i\), the above equation implies that \(i\)'s equilibrium effort in the first period is \(\tilde{R}[1 + (N-1)\beta\Delta]\). Hence it must be \(s^i_1(p)\le \tilde{R}[1 + (N-1)\beta\Delta]\), which, together with assumption \ref{assbound}, guarantees that \(\sum_{j=1}^N s^j(p)\le1\) is verified. Because \(i\) and \(p\) were generic, \ref{eqeffort1} and \ref{eqeffort2} pin down the unique equilibrium strategies of the game. 

Fix \(i\in \{1,...,N\}\) and let \(p,p^{\prime}\in [0,1]^N\) such that \(p^i = p^{\prime i}\). Using \ref{eqeffort1} and \(p^i = p^{\prime i}\),
\begin{align*}
s^i_1(p) - s^i_1(p^\prime) &= \beta\left[\mathbb{E}_{p^i}[\tilde{\omega}]\sum_{j \ne i}(e^\star_2(p^{\prime j})-e^\star_2(p^j))\right] \\
&= \beta\Delta\mathbb{E}_{p^i}[\tilde{\omega}]\sum_{j = 1}^N(p^{\prime j} - p^j) \\
&= \underbrace{N\beta\Delta\mathbb{E}_{p^i}[\tilde{\omega}]}_{>0}(\mathcal{O}(p^\prime) - \mathcal{O}(p)),
\end{align*}
which completes the proof of proposition \ref{propeffort}. \(\blacksquare\)
\paragraph{Proof of proposition \ref{propdis}}
Define the function \(\bar{e}^\star_1: [0,1]^N\to[0,1]\) such that, for any \(p\in[0,1]^N\), \(\bar{e}^\star_1(p) = \sum_{i=1}^Ns^i_1(p)\), where \(s^i(p)_1\) is  the profile of first-period effort levels played in (the most productive) equilibrium by the members of a team of type \(p\). Note that, for any \(p\in[0,1]^N\), \(\sum_{i=1}^Ne^\star_2(p^i) = Ne^\star_2(\bar{p})\), where \(\bar{p}=\mathcal{O}(p)\). Consider any two \(p,p^\prime\in[0,1]^N\) with \(\mathcal{O}(p)=\mathcal{O}(p^\prime)\), and fix \(p^\star\in[0,1]\). Note that
\begin{align}\label{eqperf}
\alpha\mathcal{Y}_{p^\star}(p) &= \bar{e}^\star_1(p)\left[\mathbb{E}_{p^\star}[\tilde{\omega}] + \beta\mathbb{E}_{p^\star}[\tilde{\omega}]\sum_{j = 1}^N(e^\star_2(p^\star)-e^\star_2(p^j))+ N\beta Var_{p^\star}(\tilde{\omega})\right] + \beta\mathbb{E}_{p^\star}[\tilde{\omega}]\sum_{i=1}^Ne^\star_2(p^i) \notag  \\
&= \bar{e}^\star_1(p)\left[\mathbb{E}_{p^\star}[\tilde{\omega}] + N\beta\mathbb{E}_{p^\star}[\tilde{\omega}](e^\star_2(p^\star)-e^\star_2(\bar{p}))+ N\beta Var_{p^\star}(\tilde{\omega})\right] + N\beta\mathbb{E}_{p^\star}[\tilde{\omega}]e^\star_2(\bar{p}),
\end{align}
for \(\bar{p}=\mathcal{O}(p)\). As a consequence,
\begin{align*}
\alpha(\mathcal{Y}_{p^\star}(p) - \mathcal{Y}_{p^\star}(p^\prime)) &= (\bar{e}^\star_1(p)-\bar{e}^\star_1(p^\prime))\left[\mathbb{E}_{p^\star}[\tilde{\omega}] + N\beta\mathbb{E}_{p^\star}[\tilde{\omega}](e^\star_2(p^\star)-e^\star_2(\bar{p}))+ N\beta Var_{p^\star}(\tilde{\omega})\right].
\end{align*}
Next, note that \(\sum_{i=1}^N\sum_{j = 1}^N\mathbb{E}_{p^i}[\tilde{\omega}](e^\star_2(p^i)-e^\star_2(p^j)) = \frac{1}{2}\sum_{i = 1}^N\sum_{j = 1}^N(e^\star_2(p^i)-e^\star_2(p^j))^2\), hence
\begin{align}
\bar{e}^\star_1(p) &= Ne^\star_2(\bar{p}) + \beta\left[\frac{1}{2}\sum_{i = 1}^N\sum_{j = 1}^N(e^\star_2(p^i)-e^\star_2(p^j))^2+ \frac{2N-1}{2}\sum_{i = 1}^NVar_{p^i}(\tilde{\omega})\right] \notag \\
&= Ne^\star_2(\bar{p}) + \beta\Delta^2\left[\frac{1}{2}\sum_{i = 1}^N\sum_{j = 1}^N(p^i-p^j)^2+ \frac{2N-1}{2}\sum_{i = 1}^Np^i(1-p^i)\right] \label{ageffort1}
\end{align}
We can rewrite \(\sum_{i = 1}^Np^i(1-p^i)\) as
\begin{align}\label{decompose}
\sum_{i = 1}^Np^i(1-p^i) &= N\bar{p}(1-\bar{p}) - N\left(\sum_{i=1}^N\frac{(p^i)^2}{N}-\bar{p}^2\right) \notag \\
&= N\left[\bar{p}(1-\bar{p})-\frac{1}{2N^2}\sum_{i = 1}^N\sum_{j = 1}^N(p^i-p^j)^2\right].
\end{align}
Plugging \ref{decompose} into \ref{ageffort1},
\begin{align*}
\bar{e}^\star_1(p) = Ne^\star_2(\bar{p}) + \frac{\beta N\Delta^2}{2}\left[(2N-1)\bar{p}(1-\bar{p}) + \frac{1}{2N^2}\sum_{i = 1}^N\sum_{j = 1}^N(p^i-p^j)^2\right],
\end{align*}
and, using the definition of \(\mathcal{D}(p)\),
\begin{equation}\label{p1eqeffort}
\bar{e}^\star_1(p) = Ne^\star_2(\bar{p}) + \frac{\beta N\Delta^2}{2}\left[(2N-1)\bar{p}(1-\bar{p}) + \frac{N-1}{2N}\mathcal{D}(p)\right].
\end{equation}
Using the above expression for \(\bar{e}^\star_1(p)\), we obtain,
\begin{equation*}
\mathcal{Y}_{p^\star}(p) - \mathcal{Y}_{p^\star}(p^\prime) = \kappa^{\mathcal{O}}_{p^\star}(\mathcal{D}(p)-\mathcal{D}(p^\prime))
\end{equation*}
for
\begin{equation*}
\kappa^{\mathcal{O}}_{p^\star} = \left[\mathbb{E}_{p^\star}[\tilde{\omega}] + N\beta\mathbb{E}_{p^\star}[\tilde{\omega}](e^\star_2(p^\star)-e^\star_2(\bar{p}))+ N\beta Var_{p^\star}(\tilde{\omega})\right]\frac{\beta\Delta^2(N-1)}{4\alpha}.
\end{equation*}
First, note that \(\kappa^{\mathcal{O}}_{p^\star}\) only depends on \(p^\star\) and \(\bar{p}\). Second, note that \(\bar{e}^\star_1(p^\star)-\bar{e}^\star_1(\bar{p})\ge-\Delta\) and that \(\mathbb{E}_{p^\star}[\tilde{\omega}](1-N\beta\Delta)> 0\) if \(N\beta\Delta <1\). \(N\beta\Delta <1\) is an immediate implication of assumption \ref{assbound}. This implies that \(\mathbb{E}_{p^\star}[\tilde{\omega}] + N\beta\mathbb{E}_{p^\star}[\tilde{\omega}](e^\star_2(p^\star)-e^\star_2(\bar{p}))>0\). Hence \(\kappa^{\mathcal{O}}_{p^\star}\) is strictly positive, which completes the proof. \(\blacksquare\)

\paragraph{Proof of proposition \ref{optimaltransport}}
Let \(N = 2\), and fix \(p^\star\in[0,1]\). Team performance is measured by \(\mathcal{Y}_{p^\star}(p^1,p^2)\). We start by showing that \(\mathcal{Y}_{p^\star}(p^1,p^2)\) is strictly submodular, that is, 
\begin{equation*}
\frac{\partial^2 \mathcal{Y}_{p^\star}(p^1,p^2)}{\partial p^1\partial p^2}<0
\end{equation*}
From equation \ref{eqperf},
\begin{align*}
\alpha\frac{\partial^2 \mathcal{Y}_{p^\star}(p^1,p^2)}{\partial p^1\partial p^2} & = \frac{\partial^2 \bar{e}^\star_1(p^1,p^2)}{\partial p^1\partial p^2} \left[\mathbb{E}_{p^\star}[\tilde{\omega}] + 2\beta\mathbb{E}_{p^\star}[\tilde{\omega}](e^\star_2(p^\star)-e^\star_2(\bar{p}))+ 2\beta Var_{p^\star}(\tilde{\omega})\right] + \\
& \quad + \frac{\partial \bar{e}^\star_1(p^1,p^2)}{\partial p^1}\frac{\partial}{\partial p^2}\left[\mathbb{E}_{p^\star}[\tilde{\omega}] + 2\beta\mathbb{E}_{p^\star}[\tilde{\omega}](e^\star_2(p^\star)-e^\star_2(\bar{p}))+ 2\beta Var_{p^\star}(\tilde{\omega})\right] + \\
& \quad + \frac{\partial \bar{e}^\star_1(p^1,p^2)}{\partial p^2}\frac{\partial}{\partial p^1}\left[\mathbb{E}_{p^\star}[\tilde{\omega}] + 2\beta\mathbb{E}_{p^\star}[\tilde{\omega}](e^\star_2(p^\star)-e^\star_2(\bar{p}))+ 2\beta Var_{p^\star}(\tilde{\omega})\right] + \\
& \quad + \bar{e}^\star_1(p^1,p^2)\frac{\partial^2}{\partial p^1\partial p^2}\left[\mathbb{E}_{p^\star}[\tilde{\omega}] + 2\beta\mathbb{E}_{p^\star}[\tilde{\omega}](e^\star_2(p^\star)-e^\star_2(\bar{p}))+ 2\beta Var_{p^\star}(\tilde{\omega})\right].
\end{align*}
Note that, from \ref{p1eqeffort}, we have 
\begin{equation*}
\bar{e}^\star_1(p^1,p^2) = 2e^\star_2(\bar{p}) + \beta\Delta^2\left[3\bar{p}(1-\bar{p}) + \frac{1}{4}(p^1-p^2)^2\right].
\end{equation*}
Hence,
\begin{gather}
\frac{\partial \bar{e}^\star_1(p^1,p^2)}{\partial p^1} = \Delta + \beta\Delta^2\left[\frac{3}{2}-p^2\right] >0 \label{partial}\\
\frac{\partial \bar{e}^\star_1(p^1,p^2)}{\partial p^2} = \Delta + \beta\Delta^2\left[\frac{3}{2}-p^1\right] >0
\end{gather}

Next, it has \(\frac{\partial}{\partial p^i}\left[\mathbb{E}_{p^\star}[\tilde{\omega}] + 2\beta\mathbb{E}_{p^\star}[\tilde{\omega}](e^\star_2(p^\star)-e^\star_2(\bar{p}))+ 2\beta Var_{p^\star}(\tilde{\omega})\right] = -\beta\Delta\mathbb{E}_{p^\star}[\tilde{\omega}]<0\) for every \(i=1,2\). Moreover, \(\frac{\partial^2}{\partial p^1\partial p^2}\left[\mathbb{E}_{p^\star}[\tilde{\omega}] + 2\beta\mathbb{E}_{p^\star}[\tilde{\omega}](e^\star_2(p^\star)-e^\star_2(\bar{p}))+ 2\beta Var_{p^\star}(\tilde{\omega})\right] = 0\), and recall that assumption \ref{assbound} implies \(\left[\mathbb{E}_{p^\star}[\tilde{\omega}] + 2\beta\mathbb{E}_{p^\star}[\tilde{\omega}](e^\star_2(p^\star)-e^\star_2(\bar{p}))+ 2\beta Var_{p^\star}(\tilde{\omega})\right]>0\). Hence,
\begin{equation*}
\frac{\partial^2 \bar{e}^\star_1(p^1,p^2)}{\partial p^1\partial p^2}<0 \implies \frac{\partial^2 \mathcal{Y}_{p^\star}(p^1,p^2)}{\partial p^1\partial p^2}<0.
\end{equation*}
But from \ref{partial} we get \(\frac{\partial^2 \bar{e}^\star_1(p^1,p^2)}{\partial p^1\partial p^2} = -\beta\Delta^2<0\), which proves the strict submodularity of \(\mathcal{Y}_{p^\star}(p^1,p^2)\). 

The submodularity of the performance function gives rise to the countermonotone coupling of proposition \ref{optimaltransport}. The following results are proven, among others, by \citet{RachevRueschendorf1998}. 
\begin{enumerate}[(i)]
\item \textbf{Existence}. They establish the existence of an optimal transport plan when \(F\) is continuous on a compact metric space and \(\mathcal{Y}_{p^\star}\) is continuous, real-valued, and bounded.
\item \textbf{Characterization}. They establish that the countermonotone coupling (negative sorting) solves the optimal transport problem when \(\mathcal{Y}_{p^\star}\) is submodular.
\item \textbf{Uniqueness}. They establish that such a solution is unique when \(\mathcal{Y}_{p^\star}\) is strictly submodular.
\end{enumerate}
All the requirements for (i), (ii), and (iii) above are met in our optimal transport problem. This observation concludes the proof of proposition \ref{optimaltransport}. \(\blacksquare\)

\paragraph{Proof of proposition \ref{workforcedis}}
Fix \(p^\star\in [0,1]\) and let \(F,G\) be two symmetric continuous distributions with support \([0,1]\). Let \(T_F\) and \(T_G\) denote the corresponding optimal transport plans of the team formation problem. By proposition \ref{optimaltransport}, \(T_F\) and \(T_G\) are countermonotone, so symmetry implies that, for each \(p\in[0,1]\), \(\mathcal{O}(p,T_G(p)) = \mathcal{O}(p,T_F(p)) = \frac{1}{2}\). By proposition \ref{propdis},
\begin{align*}
\mathcal{V}_{p^\star}(F) - \mathcal{V}_{p^\star}(G) &= \frac{1}{2}\mathcal{Y}_{p^\star}\left(\frac{1}{2},\frac{1}{2}\right) + \frac{k^{\mathcal{O}}_{p^\star}}{2}\int_{[0,1]}\mathcal{D}(p,T_F(p))dF(p) \\
&\quad - \left(\frac{1}{2}\mathcal{Y}_{p^\star}\left(\frac{1}{2},\frac{1}{2}\right) + \frac{k^{\mathcal{O}}_{p^\star}}{2}\int_{[0,1]}\mathcal{D}(p,T_G(p))dG(p)\right) \\
&= \frac{k^{\mathcal{O}}_{p^\star}}{2}\left(\int_{[0,1]}\mathcal{D}(p,T_F(p))dF(p) - \int_{[0,1]}\mathcal{D}(p,T_G(p))dG(p)\right).
\end{align*}
It remains to show that \(\int_{[0,1]}\mathcal{D}(p,T_F(p))dF(p) = 2\bar{\mathcal{D}}(F)\) and \(\int_{[0,1]}\mathcal{D}(p,T_G(p))dG(p) = 2\bar{\mathcal{D}}(G)\). We'll show the result for \(F\), but the argument is the same for \(G\). Note that by symmetry and full support of \(F\) and countermonotonicity of \(T_F\), we have \(T_F(p) = 1-p\) for each \(p\in [0,1]\). Hence \(\mathcal{D}(p,T_F(p)) = (1-2p)^2 = 4(p-\frac{1}{2})\). It follows that 
\begin{equation*}
\int_{[0,1]}\mathcal{D}(p,T_F(p))dF(p) = 4\mathbb{E}_{F}\left[\left(p-\frac{1}{2}\right)\right] = 4Var_F(p).
\end{equation*}
Next, note that, for \(p,p^\prime\sim F\) independently drawn,
\begin{align*}
\bar{\mathcal{D}}(F) &= \mathbb{E}_F[(p-p^\prime)^2] \\
&= \mathbb{E}_F\left[\left(p-\frac{1}{2} + \frac{1}{2} - p^\prime\right)^2\right] \\
&= \mathbb{E}_F\left[\left(p-\frac{1}{2}\right)^2\right] + \mathbb{E}_F\left[\left(\frac{1}{2} - p^\prime\right)^2\right] \\
&= 2Var_F(p),
\end{align*}
so that \(\frac{1}{2}\int_{[0,1]}\mathcal{D}(p,T_F(p))dF(p) = \bar{\mathcal{D}}(F)\), and \(\frac{1}{2}\int_{[0,1]}\mathcal{D}(p,T_G(p))dG(p) = \bar{\mathcal{D}}(G)\). Hence,
\begin{align*}
\mathcal{V}_{p^\star}(F) - \mathcal{V}_{p^\star}(G) &= \frac{k^{\mathcal{O}}_{p^\star}}{2}\left(\int_{[0,1]}\mathcal{D}(p,T_F(p))dF(p) - \int_{[0,1]}\mathcal{D}(p,T_G(p))dG(p)\right) \\
&= k^{\mathcal{O}}_{p^\star}(\bar{\mathcal{D}}(F)-\bar{\mathcal{D}}(G)),
\end{align*}
which proves the proposition. \(\blacksquare\)

\paragraph{Proof of proposition \ref{hdis}} 
The proof consists of two parts. First, we show that for any \(p\in [0,1]\) and \(p^\star = (p,p)\), it has 
\begin{equation*}
((1,1), (1,1))\in\arg\max_{p^L\in L}\mathcal{Y}_{p^{\star}}(p^L),
\end{equation*}
that is, the like-minded team performing the best is one where each team member is optimistic about all technologies, pinned down by prior beliefs \(p^O = ((1,1),(1,1))\). Second, for \(p^\star = (p,p)\) and \(p^C\in C\), we compare \(\mathcal{Y}_{p^\star}(p^C)\) and \(\mathcal{Y}_{p^\star}(p^O)\) as we let \(p\) range in \([0,1]\), and show that the former quantity excedes the latter for \(p\) large enough.

We start by proving the following lemma.

\begin{lemma}\label{lemma1}
Let, \(x\in [0,1], p^\star = (x,x)\) and \(p^O = ((1,1),(1,1))\). Then \(\mathcal{Y}_{p^\star}(p^O)\ge\mathcal{Y}_{p^\star}(p^L)\) for each \(p^L\in L\setminus\{p^O\}\). In particular, if \(x < 1\), then \(\mathcal{Y}_{p^\star}(p^O)>\mathcal{Y}_{p^\star}(p^L)\) for each \(p^L\in L\setminus\{p^O\}\).
\end{lemma}
\textit{Proof of Lemma \ref{lemma1}}. Fix belief \((p,q)\in [0,1]^2\) and let \(p^L\in L\) be such that \(p^L = ((p,q),(p,q))\). Without loss of generality, let \(p\ge q\). Let \(e_{k1} = e^1_{k1} + e^2_{k1}\) denote the aggregate effort level invested in period \(t=1\) by the two players in technology \(k\in\{A,B\}\). For fixed levels \(e_{A1}, e_{B1}, e_1\in [0,1)\) such that \(e_{A1} + e_{B1} = e_1\), when players follow equilibrium play in \(t=2\), it must be that 
\begin{align*}
\mathbb{E}_{p^\star}[y|e_{A1}, e_{B1}, p, q] &= (xR-xr+r)e_1 + 2\alpha\beta\{e_{A1}e_{B1}[x(2-x)R^2 + (1-x)^2r^2] \\
& + e_{A1}(1-e_{B1})[xR^2 + (1-x)(xR - xr + r)(qR - qr + r)] \\
& + e_{B1}(1-e_{A1})[xR^2 + (1-x)(xR - xr + r)(pR - pr + r)] \\
& + (1-e_{A1})(1-e_{A2})(xR - xr + r)(pR - pr +r)\}.
\end{align*}
Note that the expression above increases strictly in \(p\). It is also increasing in \(q\), strictly so if \(x<1\). We consider two cases separately. If \(x = 1\), then 
\begin{equation*}
\mathbb{E}_{p^\star}[y|e_{A1}, e_{B1}, p, q] \le G(e_{A1}, e_{B1})\equiv Re_1 + 2\alpha\beta R^2,
\end{equation*}
with the upper bound \(G(e_{A1}, e_{B1})\) achieved if and only if \(p=1\).
If \(x<1\), then, using the fact that \(e_1 = e_{A1} + e_{B1}\), we find the following upper bound for \(\mathbb{E}_{p^\star}[y|e_{A1}, e_{B1}, p, q]\),
\begin{align*}
H(e_1,e_{A1}) &\equiv (xR - xr+r)e_1 + 2\alpha\beta\{e_{A1}(e_1-e_{A1})[x(2-x)R^2 + (1-x)^2r^2] \\
& + e_{A1}(1 + e_{A1} - e_1)[xR^2 + (1-x)(xR - xr + r)R] \\
& + (e_1 - e_{A1})(1-e_{A1})[xR^2 + (1-x)(xR - xr + r)R] \\
& + (1-e_{A1})(1+e_{A1}-e_1)(xR - xr + r)R\},
\end{align*}
which is attained at \(p=q=1\). \(H(e_1,e_{A1})\) has the following properties. First,
\begin{align*}
\frac{\partial H(e_1,e_{A1})}{\partial e_1} &= (xR - xr +r) + 2\alpha\beta\{e_{A1}(1-x)[xR^2 + (1-x)r^2 - (xR - xr + r)R] \\
&+ (1-e_{A1})x[R^2 -(xR - xr + r)R]\} \ge (xR - xr + r)(1-2R)>0,
\end{align*}
where the last inequality holds by assumption \ref{assbound}. Second, it is a quadratic in \(e_{A1}\). In particular, 
\begin{align*}
\frac{\partial H(e_1,e_1^1)}{\partial e_{A1}} &= 2\alpha\beta\{(1-2e_{A1})[x(2-x)R^2 + (1-x)^2r^2] \\
& + 2(2e_{A1} - e_1)[xR^2 + (1-x)(xR- xr + r)R] \\
& + (e_1-2e_{A1})(xR - xr + r)R\}.
\end{align*}
and,
\begin{align*}
\frac{\partial^2 H(e_1,e_1^1)}{\partial e^1_1\partial e^1_1} &= 2\alpha\beta\{-2[x(2-x)R^2 + (1-x)^2r^2] \\
& + 4[xR^2 + (1-x)(xR - xr + r)R] - 2(xR - xr + r)R\} \\
& = 4\alpha\beta\{x^2R^2 - (1-x)^2r^2 + (1-2x)(xR -xr + r)R\} \\
& = 4\alpha\beta (R-r)(1-x)[x(R-r) +r] > 0
\end{align*}
Since \(H\) is a strictly convex quadratic in \(e_{A1}\), it has two candidate maximizers \(e_{A1} = 0\) and \(e_{A1} = e_1\). It is easy to check that 
\begin{align*}
H(e_1,e_1) = H(e_1,0) & = (xR - xr +r)e_1 + 2\alpha\beta R\{e_1[xR \\
& + (1-x)(xR - xr + r)] + (1-e_1)(xR - xr + r)\}.
\end{align*}
We have shown that when \(e_{A1}\) and \(e_{B1}\) are exogenously determined, the expected output of the team is maximal at \(p=q=1\) and only depends on first-period effort levels though the aggregate effort \(e_1 = e_{A1} + e_{B1}\). If \(x = 1\) the way effort is allocated across technologies does not matter. If \(x<1\), then the maximum of \(H\) given \(e_1\) is achieved if all \(e_1\) is allocated to one technology --- regardless of whether the technology is \(A\) or \(B\). In addition, the maximal team expected output is strictly increasing in \(e_1\in [0,1]\), and, for any \(e_1\in [0,1]\), if \(x<1\) then \(p=q=1\) is a necessary condition to achieve the maximal level of expected output given \(e_1\). 

To finish the proof of lemma \ref{lemma1}, it is sufficient to show that, when we let players optimize first-period effort and technology choices consistently with \(p^L\) --- i.e., when we consider equilibrium behavior at \(t = 1\) --- a team with \(p=q=1\) achieves the highest possible equilibrium first-period effort \(e_1\) among teams in \(L\), and there exists a (most-productive) equilibrium such that both members of such a team allocate effort to the same technology in \(t=1\).

In order to show that \(e_1\) is maximized when \(p=q=1\), we start by showing that the claim is true when players work on the same technology in period \(t=1\). Without loss of generality, suppose that both players use technology \(A\) in the first period. From the first order condition, for \(i=1,2\), 
\begin{equation}\label{sametech}
e^i_1(p,q) = p\Delta + \tilde{r} + \frac{3\beta}{2}[pR^2 + (1-p)(q\Delta + \tilde{r})^2 - (\max\{p,q\}\Delta + \tilde{r})^2]
\end{equation}
and, therefore,
\begin{equation*}
e^i_1(1,1) = \tilde{R}.
\end{equation*}
Note that, as per equation \ref{sametech}, \(e^i_1(p,q)\) is continuous and strictly increasing in \(p\), hence \(p = 1\) is required for it to be at a maximum. Additionally, \(e^i_1(1,q)\) does not depend on \(q\), so \(p=q=1\) maximizes first-period effort, leading to an effort level of \(e^i_1(1,1) = \tilde{R}\). We now show that if \(p=q=1\), then there exist an equilibrium such that both players work on the same technology in the first period. Without loss of generality, assume that such technology is \(A\). By working on \(A\), both players will exert an effort level of \(\tilde{R}\) in the first period. Second, at \(t=1\), they assign unit probability to the posterior belief profile \(\pi = ((1,1),(1,1))\). In any equilibrium of the game such a belief is associated to a second-period effort level of \(\tilde{R}\), giving both players a (subjective) expected second-period payoff \(\frac{3}{2}\tilde{R}^2\), regardless of second-period technology choices. Hence, each player's (subjective) expected payoff from operating \(A\) in the first period is \(\frac{3}{2}\tilde{R}^2(1+\beta)\). Can a player benefit from a deviation to using technology \(B\)? Note that the payoff that they expect to accrue in period \(t=2\) is still \(\frac{3}{2}\tilde{R}^2\), because the unique posterior belief profile that is consistent with a prior \(p^O\) is \(\pi = p^O = ((1,1),(1,1))\). So what matters is if by deviating, the player expects a first-period payoff strictly larger than \(\frac{3}{2}\tilde{R}^2\). Note that the expected instantaneous payoff when deviating to technology \(B\) is obtained by maximizing the expectation of \(u(y_1,e)\) given that the coworker invests \(\tilde{R}\) in \(B\). Such a payoff is the value of 
\begin{equation*}
\max_{e\in[0,1]} \left[2\tilde{R}e\tilde{R} + \tilde{R}e(1-\tilde{R}) + \tilde{R}(1-e)\tilde{R} -\frac{1}{2}e^2\right]
\end{equation*}
which is solved by \(e=\tilde{R}\), yielding a value of \(\frac{3}{2}\tilde{R}^2\), so that the deviation is not profitable. Note that the team effort solution \((\tilde{R},\tilde{R})\) satisfies \(e_1 = 2\tilde{R}<1\) by assumption \ref{assbound}, so that no boundary is reached.

To conclude the proof of lemma \ref{lemma1}, we need to show that if players work on different technologies at \(t=1\), the aggregate first-period effort \(e_1\) is no larger than \(2\tilde{R}\), so that indeed \(p^O\) is the like-minded team type that maximizes \(e_1\). Without loss of generality, assume that \(p\ge q\), and that at \(t=1\), player \(1\) and \(2\) work with \(A\) and \(B\) respectively. 

The first-period effort first-order conditions of the two players lead to the following best response functions,

\begin{align}
e^1_1(e^2_1) &= A^1 + B^1e^2_1 \label{foc1} \\
e^2_1(e^1_1) &= A^2 + B^2e^1_1 \label{foc2}
\end{align}
for
\begin{align*}
A^1
&\equiv p \Delta + \tilde{r} 
  + \frac{3}{2}\beta[
      p \tilde{R}^2 
      + (1-p)(q\Delta + \tilde{r})^2 
      - (p\Delta + \tilde{r})^2
    ], \\
B^1
&\equiv \frac{3}{2}\beta[
      (1-q)\Delta^2 p(1-p)
      - p\tilde{R}^2
      - (1-p)(q\Delta + \tilde{r})^2
      + (p\Delta + \tilde{r})^2
    ], \\
A^2
&\equiv q\Delta + \tilde{r}
  + \frac{3}{2}\beta[ q\tilde{R}^2
      + (1-q)(p\Delta + \tilde{r})^2
      - (p\Delta + \tilde{r})^2
    ], \\
B^2
&\equiv \frac{3}{2}\beta[
      (1-p)\Delta^2 q(1-q)
      - q\tilde{R}^2
      - (1-q)(p\Delta + \tilde{r})^2
      + (p\Delta + \tilde{r})^2
    ].
\end{align*}
In addition, simple algebra shows that \(B^1 - B^2 = 0\). Let \(B\equiv B^1\). Taking the system of conditions \ref{foc1} and \ref{foc2} and adding together the optimal effort levels, we get the following: 
\begin{align*}
e^1_1(p,q) &= \frac{A^1+BA^2}{1-(B)^2} \\
e^2_1(p,q) & = \frac{A^2+BA^1}{1-(B)^2}
\end{align*}
which is well defined as \(B<1\) due to assumption \ref{assbound}.

Next, we show that \(e^i_1(p,q)\le \tilde{R}\) for \(i=1,2\), as summarized in the next remark.

\begin{remark}\label{remark1} The following two inequalities hold.
\begin{gather}
\frac{A^1+BA^2}{1-(B)^2} \le \tilde{R} \label{ineq1} \\
\frac{A^2+BA^1}{1-(B)^2} \le \tilde{R}.\label{ineq2}
\end{gather}
\end{remark}

The remark follows from the comparative statics results by \citet{MilRob90}. To apply those results, we focus on the reduced-form effort-choice game played in period \(t=1\) when player \(1\) works on technology \(A\) and player \(2\) works on technology \(B\). Let \(\eta = -e^2_1\) and, for each \(i\), define \(v_i(e^1_1,\eta;p,q)\) to be \(i\)'s expectation of her total payoff in the game when the team has a common prior \((p,q)\), first-period effort levels are given by \((e^1_1, \eta)\) and second-period effort levels follow equilibrium play. Note that \(B = -\frac{3}{2}\beta(1-p)q\left[\Delta^2(p+q)+2r\Delta\right]\in (-1,0]\). Hence \(\frac{\partial^2v_i(e^1_1,\eta;p,q)}{\partial e^1_1\partial\eta} = -B\ge0\) for \(i=1,2\), and the game is supermodular. We now show that \(\frac{\partial^2v_1(e^1_1,\eta;p,q)}{\partial e^1_1\partial p}\ge0\) and \(\frac{\partial^2v_2(e^1_1,\eta;p,q)}{\partial \eta \partial p}\ge0\). Note that \(\frac{\partial^2v_1(e^1_1,\eta;p,q)}{\partial e^1_1\partial p} = \frac{\partial A^1}{\partial p} + \gamma \frac{\partial B}{\partial p}\) for some \(\gamma\in [0,1]\), and \(\frac{\partial^2v_2(e^1_1,\eta;p,q)}{\partial \eta\partial p} = -\frac{\partial A^2}{\partial p} - \delta \frac{\partial B}{\partial p}\), for some \(\delta\in[0,1]\). Under our parameter restrictions --- in particular, by assumption \ref{assbound} --- we have that \(\frac{\partial A^1}{\partial p}>0\) and \(\frac{\partial A^1}{\partial p} + \frac{\partial B}{\partial p}>0\), implying \(\frac{\partial^2v_1(e^1_1,\eta;p,q)}{\partial e^1_1\partial p} > 0\). Additionally, we have \(\frac{\partial A^2}{\partial p}\le 0\) and \(\frac{\partial A^2}{\partial p} + \frac{\partial B}{\partial p}\le 0\), implying \(\frac{\partial^2v_2(e^1_1,\eta;p,q)}{\partial \eta\partial p} \ge 0\). Given that \(\frac{\partial^2v_1(e^1_1,\eta;p,q)}{\partial e^1_1\partial p}\ge0\) and \(\frac{\partial^2v_2(e^1_1,\eta;p,q)}{\partial \eta \partial p}\ge0\), theorem 6 by \citet{MilRob90} implies that the equilibrium level of \(e^1_1(p,q)\) is non-decreasing in \(p\). Hence, for every \(p\in [0,1]\) and \(q\in [0,p]\), we have that player \(1\)'s effort level satisfies \(e^1_1(p,q)\le e^1_1(1,q) = \tilde{R}\), which proves inequality \ref{ineq1}. To prove \ref{ineq2} we show that if \(q\le p\) then \(e^2_1(p,q)\le e^1_1(p,q)\). Note that, using \ref{foc1} and \ref{foc2}, we get \(e^1_1(p,q) - e^2_1(p,q) \gtreqless 0\) if \(A^1 - A^2 \gtreqless 0\). But simple algebra yields \(A^1 - A^2 = (p-q)\Delta\left[1+\frac{3}{2}\beta(1-p)\Delta(1-q)\right] \ge 0\), where the inequality follows from \(0<q\le p\). Hence  \(e^2_1(p,q)\le e^1_1(p,q)\le \tilde{R}\), which proves inequality \ref{ineq2}.

%First, note that simple algebra, combined with assumption \ref{assbound}, yields $|B|<1$, and hence $1-(B)^2>0$. Hence, it is sufficient to prove that \(N\equiv \tilde{R}[1-(B)^2]-(A^1+BA^2)\ge 0\) and \(M\equiv \tilde{R}[1-(B)^2]-(A^2+BA^1)\ge 0\). We start with the first inequality. The expression $N$ is a polynomial in
%$(p,q,\tilde R,\tilde r,\beta)$. Consider the closed parameter set defined by
%\begin{gather*}
%p\ge 0, \quad 1-p\ge 0, \\
%q\ge 0,\quad p-q\ge 0, \\
%\tilde r\ge 0,\quad \tilde R-\tilde r\ge 0, \\
%\beta\ge 0, \quad 1-\beta\ge 0, \\
%\tfrac14-\tilde R\ge 0,
%\end{gather*}
%where the last constraint is simply a restatement of assumption \ref{assbound} in our context.
%Using a sum-of-squares (SOS) relaxation\footnote{The relaxation is computed with \texttt{SumOfSquares.jl}
%and the \texttt{SCS} solver.} of degree $6$ we obtain an explicit identity
%\[
%4N
%=
%\nu_0 + \sum_{i=1}^9 \nu_i g_i,
%\]
%where $g_i$ are the constraint polynomials above and each $\nu_i$ is a sum of squares.
%Since $\nu_i\ge 0$ and $g_i\ge 0$ on the feasible set, the right-hand side is nonnegative,
%implying $N\ge 0$. This proves that \(N\ge 0\). \(M\ge 0\) is proved in the same way. The expression $M$ is a polynomial in
%$(p,q,\tilde R,\tilde r,\beta)$, defined on the same closed parameter set reported above. An SOS relaxation of degree $6$ returns an explicit certificate
%\[
%4M
%=
%\mu_0 + \sum_{i=1}^9 \mu_i g_i,
%\]
%with each $\mu_i$ a sum-of-squares polynomial. Therefore $M\ge 0$ on the feasible set.

Remark \ref{remark1} implies that, if an equilibrium where both players work on different technology exists, the maximal level of aggregate first-period effort exerted in the first period is \(e_1(1,1) = 2\tilde{R}\). Summarizing our findings starting from this final step of the proof, we have shown that, when we restrict attention to beliefs in \(L\), a team with initial beliefs \(p^O\) maximizes the equilibrium first-period aggregate effort, and therefore the expected first-period output evaluated at belief \(p^\star = (x,x)\). In addition, we have shown that belief profile \(p^O\) admits a most productive equilibrium where players invest this maximal effort level on the same technology, maximizing the expected second period output when evaluated at \(t=1\) according to belief profile \(p^\star\). Finally, we have shown that \(p^O\) is the unique maximizer of such expected second-period output if \(x<1\). This completes the proof of lemma \ref{lemma1}. \(\blacksquare\)

We now continue the proof of proposition \ref{hdis}. Note that for \(p^\star = (p,p)\), \(p\in[0,1]\) it has
\begin{equation*}
\alpha\mathcal{Y}_{p^\star}(p^O) = 2\tilde{R}(p\Delta+\tilde{r}) + 2\beta[\tilde{R}p(1-p)\Delta(\tilde{R}+\tilde{r} + p\Delta)+p\Delta +\tilde{r}] 
\end{equation*}
Now, fix \(p^C\in C\) and let us compute \(\alpha\mathcal{Y}_{p^\star}(p^L)\). Without loss of generality, let \(p^C = ((1,0),(0,1))\). Assume that in equilibrium player \(1\) invests effort \(e^1_1\in [0,\frac{1}{2}]\) in technology \(A\), and player \(2\) invests effort \(e^2_1\) in technology \(B\) --- we will show that this is the case in a later stage of the proof. Note that if only one technology produces \(\tilde{R}\) at \(t=1\) then in the most productive equilibrium both players use that technology at \(t=2\), even if one of them is indifferent between technologies. Hence, conditional on \(e^1_1\) and \(e^1_2\), we have
\begin{align}\label{objY}
\alpha\mathcal{Y}_{p^\star}(p^C) &= (p\Delta + \tilde{r})e_1 + \beta\bigg\{2e^1_1e^2_1\bigg[p(2-p)\tilde{R}^2 + (1-p)^2\tilde{r}^2\bigg] \notag \\
&+ e^1_1(1-e^2_1)\left[p2\tilde{R}^2 + (1-p)(p\Delta+\tilde{r})(\tilde{R}+d_1\tilde{r}) + (1-p)(1-d_1)\tilde{r}^2\right] \notag \\ 
&+ e^2_1(1-e^1_1)\left[p2\tilde{R}^2 + (1-p)(p\Delta+\tilde{r})(\tilde{R}+d_2\tilde{r}) + (1-p)(1-d_2)\tilde{r}^2\right] \notag \\
&+ 2(1-e^1_1)(1-e^2_1)(p\Delta+\tilde{r})\tilde{R}\bigg\}
\end{align}
where \(e_1 = e_1^1+e_1^2\) and \(d_i\in\{0,1\}\) is an indicator capturing the technology that \(i\) operates according to the most productive equilibrium strategy after observing that the technology she operates yields return \(\tilde{r}\) and the other technology does not yield a return. In fact, after such an event,  \(i\)'s posterior beliefs are such that she is indifferent between changing technology between the two periods (\(d_i = 1\)) or operating in the second period the same technology used in the first (\(d_i = 0\)).

%We start by proving that \(\mathcal{Y}_{(p,p)}(p^O)\) is increasing \(p\). Note that 
%\begin{equation*}
%\alpha\frac{\partial \mathcal{Y}_{(p,p)}(p^O)}{\partial{p}} \ge 2\Delta(\tilde{R}+\beta)+2\beta\tilde{R}\Delta(\tilde{R}+\tilde{r} + p\Delta)(1-2p)>2\beta\Delta[1-\tilde{R}(\tilde{R}+\tilde{r})]>0,
%\end{equation*}
%where the last inequality holds by assumption \ref{assbound}. Second, note that,
%\begin{equation*}
%\alpha\frac{\partial \mathcal{Y}_{(p,p)}(p^C)}{\partial{p}} \ge \Delta e_1+4\beta e^1_1e^1_2(1-p)(\tilde{R}+\tilde{r})\Delta + 2(1-e^1_1)(1-e^2_1)\Delta\tilde{R}>0.
%\end{equation*}
%Hence, both \(\mathcal{Y}_{(p,p)}(p^O)\) and \(\mathcal{Y}_{(p,p)}(p^C)\) are strictly increasing in \(p\). 
Before proceeding, it is useful to derive the pairs first-period equilibrium effort levels \(e^1_1\) and \(e^2_1\) of a team with beliefs \(p^C\). In any most productive equilibrium, if exactly one technology produces \(\tilde{R}\) at \(t=1\), then both players operate that technology in period \(t=2\) --- the productive history makes one of the two players indifferent between the two technologies at \(t=2\), and strategies that dictate a switch to the successful technology motivate the other player to exert more effort in the first period. We need to consider three cases, based on what technology player \(j\in\{1,2\}\) uses at \(t=2\) after observing that the technology she used at \(t=1\) yielded a return of \(\tilde{r}\) and the other technology yielded no return (note that in such a situation \(j\) is indifferent between \(A\) and \(B\) at \(t=2\)). 

\textit{Case 1: \(d_1=d_2=1\)}. The first-order condition identifying the first-period effort leads to
\begin{equation}\label{1}
\tilde{e}^i_1 = \tilde{R} + \beta[e^{-i}_1\tilde{R}\Delta + (1-e^{-i}_1)\tilde{R}\Delta] = \tilde{R}(1+\beta\Delta), \quad i = 1,2.
\end{equation}
Note that \(\tilde{e}^1_1+\tilde{e}^2_1<1\) by assumption \ref{assbound}. To show that this case represents an equilibrium, we need to verify that no player has an incentive to operate a different technology at \(t=1\). Assume that player \(2\) invests effort \(\tilde{e}^2_1\) in technology \(B\) and that player \(1\) benefits from a deviation to investing \(x\in[0,1-\tilde{e}^2_1]\) in technology \(B\) instead of \(\tilde{e}^1_1\) in technology \(A\). According to player \(1\)'s beliefs, the (constant) expected marginal benefit of effort in technology \(B\) is \(\tilde{r}\), which is strictly less than \(\tilde{R}(1 + \beta\Delta)\), the (constant) marginal benefit of effort in technology \(A\) pinned down by the right-hand side of \ref{1}. But this means that, since effort costs are the same across technology, player \(1\) would benefit from investing \(x\) in \(A\) instead of \(B\), which is a contradiction since \(\tilde{e}^1_1\) is optimal when player \(1\) works on technology \(A\). The proof that player \(2\) has no deviation incentives is analogous.

\textit{Case 2: \(d_1=d_2=0\)}. The equilibrium first-period effort levels are pinned down by the system of the following two first-order conditions,

\begin{align}
e_1^1 &= \tilde{R} + \beta[e^{2}_1(\tilde{R} + \tilde{r})\Delta + (1-e^{2}_1)\tilde{R}\Delta] = \tilde{R}(1 + \beta\Delta) + e^2_1\beta\tilde{r}\Delta \label{deviation2} \\
e^2_1 &= \tilde{R} + \beta[e^{1}_1(\tilde{R}+r)\Delta + (1-e^{1}_1)\tilde{R}\Delta] = \tilde{R}(1 + \beta\Delta) + e^1_1\beta\tilde{r}\Delta
\end{align}
with solution
\begin{equation}\label{2}
\hat{e}^i_1 = \frac{\tilde{R}(1 + \beta\Delta)}{1-\beta\tilde{r}\Delta} = \tilde{R}\left(1 + \frac{1+\tilde{r}}{1-\beta \tilde{r}\Delta}\beta\Delta\right).
\end{equation}
Note that \(\hat{e}^1_1+\hat{e}^2_1<1\) by assumption \ref{assbound}. To show that this case represents an equilibrium, we need to verify that no player has an incentive to operate a different technology at \(t=1\). Assume that player \(2\) invests effort \(\hat{e}^2_1\) in technology \(B\) and that player \(1\) benefits from a deviation to investing \(x\in[0,1-\hat{e}^2_1]\) in technology \(B\) instead of \(\hat{e}^1_1\) in technology \(A\). According to player \(1\)'s beliefs, the (constant) expected marginal benefit of effort in technology \(B\) is \(\tilde{r} - \beta\tilde{r}\Delta\), which is strictly less than \(\tilde{R}(1 + \beta\Delta) + \hat{e}^2_1\beta\tilde{r}\Delta\), the (constant) marginal benefit of effort in technology \(A\) pinned down by the right-hand side of \ref{deviation2}. But this means that, since effort costs are the same across technologies, player \(1\) would benefit from investing \(x\) in \(A\) instead of \(B\), which is a contradiction since \(\hat{e}^1_1\) is optimal when player \(1\) works on technology \(A\). The proof that player \(2\) has no deviation incentives is analogous.

\textit{Case 3: \(d_1\ne d_2\)}. Without loss of generality, assume \(d_2 =1\). Then from the system of first-order conditions we obtain the following first-period equilibrium effort levels,

\begin{align}
\breve{e}_1^1 &= \tilde{R}(1+\beta\Delta) \label{3.1} \\
\breve{e}^2_1 &= \tilde{R}(1+\beta\Delta)(1+\beta\tilde{r}\Delta) \label{3.2}
\end{align}
Also in this case, \(\breve{e}^1_1+\breve{e}^2_1<1\) by assumption \ref{assbound}. To show that this case represents an equilibrium, we need to verify that no player has an incentive to operate a different technology at \(t=1\). Assume that player \(2\) invests effort \(\breve{e}^2_1\) in technology \(B\) and that player \(1\) benefits from a deviation to investing \(x\in[0,1-\breve{e}^2_1]\) in technology \(B\) instead of \(\breve{e}^1_1\) in technology \(A\). According to player \(1\)'s beliefs, the (constant) expected marginal benefit of effort in technology \(B\) is \(\tilde{r}\), which is strictly less than the (constant) marginal benefit of effort in technology \(A\) pinned down by the right-hand side of \ref{3.1}. But this means that, since effort costs are the same across technology, player \(1\) would benefit from investing \(x\) in \(A\) instead of \(B\), which is a contradiction since \(\breve{e}^1_1\) is optimal when player \(1\) works on technology \(A\). Now, assume that player \(1\) invests effort \(\breve{e}^1_1\) in technology \(A\) and that player \(2\) benefits from a deviation to investing \(x\in[0,1-\breve{e}^1_1]\) in technology \(A\) instead of \(\breve{e}^2_1\) in technology \(B\). According to player \(2\)'s beliefs, the (constant) expected marginal benefit of effort in technology \(A\) is \(\tilde{r} - \beta\tilde{r}\Delta\), which is strictly less than the (constant) marginal benefit of effort in technology \(B\) pinned down by the right-hand side of \ref{3.2}. But this means that, since effort costs are the same across technologies, player \(2\) would benefit from investing \(x\) in \(B\) instead of \(A\), which is a contradiction since \(\breve{e}^2_1\) is optimal when player \(2\) works on technology \(B\).

We now show \(\mathcal{Y}_{(0,0)}(p^C)\gtreqless\mathcal{Y}_{(0,0)}(p^O)\) if \(\beta\lesseqgtr\frac{\tilde{r}}{\Delta}\). Assume \(p^\star = (0,0)\). In this case we have \(\alpha\mathcal{Y}_{(0,0)}(p^O) = 2\tilde{R}\tilde{r}(1+\beta)\).
Let us now find \(\mathcal{Y}_{(0,0)}(p^C)\). From equation \ref{objY}, we have 
\begin{align}\label{case2}
\alpha\mathcal{Y}_{p^\star}(p^C) &= \tilde{r}e_1 + \beta[2e^1_1e^2_1\tilde{r}^2 + e^1_1(1-e^2_1)\tilde{r}(\tilde{R}+\tilde{r}) + e^2_1(1-e^1_1)\tilde{r}(\tilde{R}+\tilde{r}) \notag \\
&+ 2(1-e^1_1)(1-e^2_1)\tilde{r}\tilde{R}]
\end{align}
Consider \textit{Case 1} first. In this case, we have \(d_1=d_2=1\). Plugging effort levels \((\tilde{e}^1_1,\tilde{e}^1_1)\) in \ref{case2}, we obtain 
\begin{align*}
\alpha\tilde{\mathcal{Y}}_{(0,0)}(p^C) &= 2\tilde{r}\tilde{e}^1_1 + 2\beta[\tilde{e}_1^1\tilde{r}^2 + (1-\tilde{e}_1^1)\tilde{R}\tilde{r}] \\
&= 2\tilde{r}\tilde{e}^1_1(1-\beta\Delta) + 2\beta\tilde{R}\tilde{r} \\
&= 2\tilde{R}\tilde{r}[1+\beta(1 - \beta\Delta^2)] < 2\tilde{R}\tilde{r}(1+\beta) = \alpha\mathcal{Y}_{(0,0)}(p^O).
\end{align*}
Next, consider \textit{Case 2}. In this case, we have \(d_1=d_2=0\). Plugging effort levels \((\hat{e}^1_1,\hat{e}^1_1)\) in \ref{case2}, we obtain, 
\begin{align*}
\alpha\hat{\mathcal{Y}}_{(0,0)}(p^C) &= 2\hat{r}\hat{e}^1_1 + 2\beta[\hat{e}_1^1\tilde{r}^2 + (1-\hat{e}_1^1)\tilde{R}\tilde{r}] \\
&= 2\tilde{r}\hat{e}^1_1(1-\beta\Delta) + 2\beta\hat{R}\tilde{r} \\
&= 2\tilde{R}\tilde{r}\left(\frac{1 - \beta^2\Delta^2}{1-\beta\tilde{r}\Delta}+\beta\right)
\end{align*}
Hence, in this case, we have 
\begin{equation*}
\hat{\mathcal{Y}}_{(0,0)}(p^C)\gtreqless \mathcal{Y}_{(0,0)}(p^O) \text{ if } \beta\lesseqgtr\frac{\tilde{r}}{\Delta}.
\end{equation*}

Finally, consider \textit{Case 3}, so that \(d_1 = 0\) and \(d_2 = 1\). First, note that we can rewrite \ref{case2} as
\begin{align*}
\alpha\mathcal{Y}_{(0,0)}(p^C) &= \tilde{r}e_1 + \beta[e_1\tilde{r}^2 + (2-e_1)\tilde{R}\tilde{r}] \\
& = e_1\tilde{r}(1-\beta\Delta) + 2\beta\tilde{R}{r}.
\end{align*}
Plugging \(e_1 = \breve{e}^1_1 + \breve{e}^2_1 = \tilde{R}(1+\beta\Delta)(2+\beta\tilde{r}\Delta)\), we get
\begin{align*}
\alpha\mathcal{\breve{Y}}_{(0,0)}(p^C) & = \tilde{R}\tilde{r}(1+\beta\Delta)(2+\beta\tilde{r}\Delta)(1-\beta\Delta) + 2\beta\tilde{R}{r} \\
& = \tilde{R}\tilde{r}[(1-\beta^2\Delta^2)(2+\beta\tilde{r}\Delta) + 2\beta].
\end{align*}
Hence we have that
\begin{align*}
\hat{\mathcal{Y}}_{(0,0)}(p^C)>\mathcal{\breve{Y}}_{(0,0)}(p^C) &\iff \frac{1}{1-\beta\tilde{r}\Delta}>1+\frac{\beta\tilde{r}\Delta}{2} \\
&\iff 1 > 1-(\beta\tilde{r}\Delta)^2 - \frac{\beta\tilde{r}\Delta}{2}(1-\beta\tilde{r}\Delta),
\end{align*}
which always holds true because of assumption \ref{assbound}. 

By definition, it has \(\mathcal{Y}_{(0,0)}(p^O) = \max\{\tilde{\mathcal{Y}}_{(0,0)}(p^C), \hat{\mathcal{Y}}_{(0,0)}(p^C), \breve{\mathcal{Y}}_{(0,0)}(p^C)\). Since \(\mathcal{Y}_{(0,0)}(p^O)>\tilde{\mathcal{Y}}_{(0,0)}(p^C)\), \(\hat{\mathcal{Y}}_{(0,0)}(p^C)>\mathcal{\breve{Y}}_{p^\star}(p^C)\), and \(\hat{\mathcal{Y}}_{(0,0)}(p^C)\gtreqless \mathcal{Y}_{(0,0)}(p^O)\) if \(\beta\lesseqgtr\frac{\tilde{r}}{\Delta}\), we conclude that 
\begin{equation*}
\mathcal{Y}_{(0,0)}(p^C)\gtreqless \mathcal{Y}_{(0,0)}(p^O) \text{ if } \beta\lesseqgtr\frac{\tilde{r}}{\Delta}.
\end{equation*}
Next, note that it must be \(\mathcal{Y}_{(1,1)}(p^C)> \mathcal{Y}_{(1,1)}(p^O)\). In fact plugging \(p=1\) in \ref{objY}, we have 
\begin{equation*}
\alpha\mathcal{Y}_{(1,1)}(p^C) = e_1\tilde{R} + 2\beta\tilde{R}^2
\end{equation*}
where \(e_1\) is the first-period aggregate effort \(e_1 = e^1_1+e^2_1\) in the most productive equilibrium given beliefs \((p^\star,p^C)\). Hence
\begin{equation*}  
\alpha(\mathcal{Y}_{(1,1)}(p^C)-\mathcal{Y}_{(1,1)}(p^O)) = (e_1-2\tilde{R})\tilde{R}>0
\end{equation*}
where the inequality follows from the fact that \(\min\{\tilde{e}_1^1+\tilde{e}_1^2,\hat{e}_1^1+\hat{e}_1^2,\breve{e}_1^1+\breve{e}_1^2\}> 2\tilde{R}\).

Let \(\tilde{\delta}(p)\equiv \tilde{\mathcal{Y}}_{(p,p)}(p^C)-\mathcal{Y}_{(p,p)}(p^O)\), \(\hat{\delta}(p)\equiv \hat{\mathcal{Y}}_{(p,p)}(p^C)-\mathcal{Y}_{(p,p)}(p^O)\), and \(\breve{\delta}(p)\equiv \breve{\mathcal{Y}}_{(p,p)}(p^C)-\mathcal{Y}_{(p,p)}(p^O)\). To finish the proof, it is sufficient to inspect \ref{objY} and note that, for any \(e^1_1,e^2_1\in(0,1)\) and any \(d_1,d_2\in\{0,1\}\), the right-hand side is a concave quadratic in \(p\), while \(\alpha\mathcal{Y}_{(p,p)}(p^O) = 2(p\Delta+\tilde{r})\tilde{R}(1+\beta)\) is affine in \(p\). Hence \(\tilde{\delta}, \hat{\delta}\) and \(\breve{\delta}\) are all concave quadratic in \(p\). This has the following three implications. First, if \(\beta < \frac{\tilde{r}}{\Delta}\), then we have \(\hat{\delta}(0)>0\) and \(\hat{\delta}(1)>0\). Since \(\hat{\delta}\) is a concave quadratic in \(p\), we have \(\hat{\delta}(p)>0\) for all \(p\in[0,1]\). This implies that when \(\beta < \frac{\tilde{r}}{\Delta}\), \(\mathcal{Y}_{(p,p)}(p^C)-\mathcal{Y}_{(p,p)}(p^O)>0\) for all \(p\in[0,1]\). Second, if \(\beta = \frac{\tilde{r}}{\Delta}\), then \(\max\{\tilde{\delta}(0), \hat{\delta}(0),\breve{\delta}(0)\} = 0\) and \(\min\{\tilde{\delta}(1), \hat{\delta}(1),\breve{\delta}(1)\}  > 0\). These relations imply that, since the three functions are concave quadratic in \(p\), there must exist three unique thresholds \(\tilde{p},\hat{p},\breve{p}\in [0,1)\) associated to \(\tilde{\delta}, \hat{\delta}\) and \(\breve{\delta}\), respectively, such that \(\delta\in \{\tilde{\delta}, \hat{\delta},\breve{\delta}\}\) is strictly positive (negative) if \(p\) is strictly greater (less) than the respective threshold \(p_{\delta}\in\{\tilde{p},\hat{p},\breve{p}\}\), and \(\delta(p_\delta) = 0\). In particular, we have \(\hat{p}=0\) and \(\tilde{p},\breve{p}>0\). This implies that, if \(\beta = \frac{\tilde{r}}{\Delta}\), then \(\mathcal{Y}_{(p,p)}(p^C)-\mathcal{Y}_{(p,p)}(p^O) = 0\) if \(p = 0\) and \(\mathcal{Y}_{(p,p)}(p^C)-\mathcal{Y}_{(p,p)}(p^O) > 0\) if \(p > 0\). Finally, if \(\beta > \frac{\tilde{r}}{\Delta}\), then \(\max\{\tilde{\delta}(0), \hat{\delta}(0),\breve{\delta}(0)\} < 0\) and \(\min\{\tilde{\delta}(1), \hat{\delta}(1),\breve{\delta}(1)\} > 0\). Hence, since the three functions are concave quadratic in \(p\), there must exist three unique thresholds \(\tilde{p},\hat{p},\breve{p}\in (0,1)\) associated to \(\tilde{\delta}, \hat{\delta}\) and \(\breve{\delta}\), respectively, such that \(\delta\in \{\tilde{\delta}, \hat{\delta},\breve{\delta}\}\) is strictly positive (negative) if \(p\) is strictly greater (less) than the respective threshold \(p_{\delta}\in\{\tilde{p},\hat{p},\breve{p}\}\), and \(\delta(p_\delta) = 0\). Let \(\bar{p} \equiv \min\{\tilde{p},\hat{p},\breve{p}\}\). The above implies \(\max\{\tilde{\delta}(0), \hat{\delta}(0),\breve{\delta}(0)\} \gtreqless 0\) if \(p\gtreqless \bar{p}\) or, equivalently, \(\mathcal{Y}_{(p,p)}(p^C)-\mathcal{Y}_{(p,p)}(p^O) \gtreqless 0\) if \(p\gtreqless \bar{p}\). \(\blacksquare\)

\paragraph{Proof of proposition \ref{welfare}} 
Let \(p^\star = (1,1)\) and \(p^C\in C\). First, we show that \(\mathcal{W}^i_{p^\star}(p^C)>\mathcal{W}^i_{p^\star}(p^O)\) for \(i=1,2\) and \(p^O = ((1,1),(1,1))\). Note that, for any \(i,t\in{1,2}\) any effort levels \(e = (e^i_t)_{i,t\in\{1,2\}}\in [0,1]^4\) and technology choices \(k=(k^i_t)_{i,t\in\{1,2\}}\in\{A,B\}^4\), the joint expected payoff of the two players is
\begin{equation*}
\hat{\mathcal{W}}_{p^\star}(e,k) = \sum_{i,t\in\{1,2\}}\mathbb{E}_{p^\star}\left[u(y_1,e^i_t)|e,k\right] = 2\tilde{R}(e_1 + e_2\beta) - \frac{1}{2}\sum_{i=1,2}(e^i_1)^2 - \frac{\beta}{2}\sum_{i=1,2}(e^i_2)^2
\end{equation*}
where \(e_t = \sum_{i\in\{1,2\}}e^i_t\). The following three observations are immediate to see. First, \(\hat{\mathcal{W}}_{p^\star}(e,k)\) does not depend on \(k\). Second, \(\hat{\mathcal{W}}_{p^\star}(e,k)\) is maximized by setting \(e^i_t = 2\tilde{R}\) for all \(i,t=\{1,2\}\), and at the optimum \(e_t = 4\tilde{R}<1\), so that assumption \ref{assbound} guarantees that in each period the probability of a productive success is well defined. Third, if we impose the constraints \(e^1_1 + e^2_1 \le b_1\) and \(e^1_2 + e^2_2 \le b_2\), with \(b_1,b_2\in[0,4\tilde{R}]\), then \(\hat{\mathcal{W}}_{p^\star}(e,k)\) is maximized at \(e^1_1=e^2_1 = \frac{b_1}{2}\) and \(e^1_2=e^2_2 = \frac{b_2}{2}\), and the value of the maximization is strictly increasing in \(b_1\) and \(b_2\) on \([0,4\tilde{R}]\). Given the above three observations, to prove that \(\mathcal{W}^i_{p^\star}(p^C)>\mathcal{W}^i_{p^\star}(p^O)\) for \(i=1,2\), it is sufficient to show that, when we compare the most productive equilibria for any team prior belief \(p^C\in C\) and any team prior belief \(p^O\) evaluated at belief \(p^\star\), then on any equilibrium path occurring with positive probability: (i) the team of type \(p^C\) exerts a strictly larger aggregate effort \(e_1\) than the team of type \(p^O\); (ii) the team of type \(p^C\) exerts a weakly larger aggregate effort \(e_2\) than the team of type \(p^O\); (iii) in both types of teams and periods, aggregate effort is equally split between players, that is \(e^1_t = e^2_t\) for \(t=1,2\). 

First, note that in the proof of proposition \ref{hdis}, we have shown that, in any equilibrium, team \(p^C\) exerts a strictly higher level of aggregate first-period effort \(e_1\) than team \(p^O\), which proves (i). Moving to (ii), note that, for any team belief \(p\in[0,1]^4\) the highest possible level of second-period effort \(e^i_2\) exerted by player \(i\) on an equilibrium path is \(\tilde{R}\) and is attained if and only if the player holds a posterior belief in \(\Pi = \{(\pi_A,\pi_B)\in [0,1]^2 : \max\{\pi_A,\pi_B\} = 1\}\); when the team is of type \(p^C\) and equilibrium-path posterior beliefs are consistent \(p^\star = (1,1)\), then for both players the only feasible posterior beliefs are elements of \(\Pi\), which proves (ii). As for (iii), we first show that in the most productive equilibrium for team \(p^C\) at belief \(p^\star = (1,1)\), the two players exert the same level of effort at \(t=1\). In the proof of proposition \ref{hdis}, we have shown that the most productive equilibrium first-period effort choices for team \(p^C\) belong to the set \(\{(\tilde{e}_1^1,\tilde{e}_1^2),(\hat{e}_1^1,\hat{e}_1^2),(\breve{e}_1^1,\breve{e}_1^2)\}\), where the three effort profiles are defined by equations \ref{1}, \ref{2}, \ref{3.1}, and \ref{3.2}. At \(p^\star = (1,1)\) the most productive equilibrium for team is the one yielding the highest level of first period aggregate effort. But from the definitions, we have \(\hat{e}_1^1+\hat{e}_1^2 >\max\{\tilde{e}_1^1+\tilde{e}_1^2,\breve{e}_1^1+\breve{e}_1^2\}\) and \(\hat{e}_1^1 = \hat{e}_1^2\). To conclude, as shown in proposition \ref{hdis}, when a team is of type \(p^O\) and \(p^\star = (1,1)\), then in any equilibrium both players exert effort \(\tilde{R}\) in both periods. This proves (iii). Given the properties of \(\hat{\mathcal{W}}_{p^\star}(e,k)\) outlined above, results (i), (ii), and (iii) imply \(\mathcal{W}^i_{p^\star}(p^C)>\mathcal{W}^i_{p^\star}(p^O)\) for \(i=1,2\).

We are left to show that \(\mathcal{W}^i_{p^\star}(p^O)\ge\mathcal{W}^i_{p^\star}(p^L)\) for each \(p^L\in L\) and \(i=1,2\). For any \(i,t\in{1,2}\) any effort levels \(e = (e^i_t)_{i,t\in\{1,2\}}\in [0,1]^4\) and technology choices \(k=(k^i_t)_{i,t\in\{1,2\}}\in\{A,B\}^4\), the expected payoff of player \(i\) evaluated at \(p^\star = (1,1)\) is
\begin{equation*}
\hat{\mathcal{W}}^i_{p^\star}(e,k) = \sum_{t\in\{1,2\}}\mathbb{E}_{p^\star}\left[u(y_1,e^i_t)|e,k\right] = \tilde{R}(e_1 + e_2\beta) - \frac{1}{2}(e^i_1)^2 - \frac{\beta}{2}(e^i_2)^2
\end{equation*}
for \(e_t = \sum_{i\in\{1,2\}}e^i_t\). This expected payoff is increasing in \(e^{-i}_1\) and \(e^{-i}_2\), and, for any fixed \(e^{-i}_1,e^{-i}_2\in[0,1-\tilde{R}]\), it reaches a maximum at \(e^{i}_1=e^{i}_2=\tilde{R}\). When players hold beliefs \(p^O\) and \(p^\star = (1,1)\), their on path equilibrium effort levels are \(e^1_1=e^2_1=e^1_2=e^2_2=\tilde{R}\). In addition, recall that for any team prior beliefs \(p\in[0,1]^4\) and any corresponding equilibrium, \(\tilde{R}\) is the maximum level of effort that can be exerted in equilibrium at \(t=2\). As a result, it is sufficient that for any equilibrium or the game and prior belief profile \(p^L\in L\), \(p^L = ((p,q),(p,q))\), players' first-period effort levels at \(p^L\) are such that \(e^i_1(p,q)\le \tilde{R}\) for \(i=1,2\) --- a final step that follows immediately from remark \ref{remark1}. This completes the proof. \(\blacksquare\)
\end{document}